\documentclass[fleqn, usenatbib]{mnras}

\usepackage{newtxtext,newtxmath}
\usepackage[T1]{fontenc}
\usepackage{ae,aecompl}
\usepackage{graphicx}	        
\usepackage{amsmath}	       
\usepackage{url}
\usepackage{microtype}
\usepackage{xspace}
\usepackage{float}
\usepackage[usenames]{color}
\usepackage{extdash}
\usepackage{makecell}

\definecolor{green}{rgb}{0.3,0.7,0.}

\definecolor{orange}{rgb}{0.075, 0.533, 0.015}

\definecolor{black}{rgb}{0, 0, 0}
\newcommand\new[1]{{\color{black} #1}}

\newcommand{\mcore}{\ensuremath{M_{\rm core}}\xspace}
\newcommand{\menv}{\ensuremath{M_{\rm env}}\xspace}
\newcommand{\xc}{\ensuremath{X_{\rm c}}\xspace}
\newcommand{\yc}{\ensuremath{Y_{\rm c}}\xspace}
\newcommand{\msun}{\ensuremath{\mathrm{M}_{\odot}}\xspace}
\newcommand{\aov}{\ensuremath{\alpha_{\rm ov}}\xspace}
\newcommand{\teff}{\ensuremath{T_{\rm eff}}\xspace}

\newcommand{\mucore}{\ensuremath{\mu_{\rm core}}\xspace}

\newcommand{\lnuc}{\ensuremath{L_{\rm nuc}}\xspace}

\newcommand{\lsun}{\ensuremath{\mathrm{L}_{\odot}}\xspace}
\newcommand{\rsun}{\ensuremath{\mathrm{R}_{\odot}}\xspace}

\newcommand{\fov}{\ensuremath{f_{\rm ov}}\xspace}

\newcommand{\logllsun}{\ensuremath{\log \mathrm{L}/\lsun}\xspace}

\newcommand{\logteff}{\ensuremath{\log T_{\rm eff}}\xspace}
\newcommand{\coremassratio}{\ensuremath{M_{\mathrm{core}}/M_{\mathrm{total}}}\xspace}

\newcommand{\mesa}{\textsc{mesa}\xspace}
\newcommand{\snap}{\textsc{snapshot}\xspace}
\newcommand{\cnoshell}{\ensuremath{{\rm CNO}_{\rm shell}}\xspace}
\newcommand{\yshell}{\ensuremath{{\rm Y}_{\rm shell}}\xspace}
\newcommand{\xenv}{\ensuremath{{\rm X}_{\rm env}}\xspace}
\newcommand{\ycore}{\ensuremath{Y_{\rm core}}\xspace}
\newcommand{\lactual}{\ensuremath{L_{\rm actual}}\xspace}
\newcommand{\muu}{\ensuremath{\mu}\xspace}
\newcommand{\muavg}{\ensuremath{\mu_{\rm avg}}\xspace}
\newcommand{\ly}{$L$\xspace}
\newcommand{\egrav}{\ensuremath{\epsilon_{\rm grav}}\xspace}

\title[Cause and Effect in Massive Star Evolution]{Numerical experiments to help understand cause and effect in massive star evolution}

\author[E. Farrell et al.]{
Eoin Farrell$^{1}$\thanks{E-mail: efarrel4@tcd.ie},
Jose H. Groh$^{1}$,
Georges Meynet$^{2}$,
and JJ Eldridge$^{3}$
\\
$^{1}$School of Physics, Trinity College Dublin, The University of Dublin, Dublin, Ireland \\
$^{2}$Geneva Observatory, University of Geneva, Chemin des Maillettes 51, 1290 Sauverny, Switzerland \\
$^{3}$Department of Physics, Private Bag 92019, University of Auckland, New Zealand
}

\pubyear{2022}

\begin{document}
\label{firstpage}
\pagerange{\pageref{firstpage}--\pageref{lastpage}}
\maketitle

\begin{abstract}
The evolution of massive stars is affected by a variety of physical processes including convection, rotation, mass loss and binary interaction.
Because these processes modify the internal chemical abundance profiles in multiple ways simultaneously, it can be challenging to determine which properties of the stellar interior are primarily driving the overall evolution.
Building on previous work, we develop a new modelling approach called \textsc{snapshot} that allows us to isolate the key features of the internal abundance profile that drive the evolution of massive stars.
Using our approach, we compute numerical stellar structure models in thermal equilibrium covering key phases of stellar evolution.
For the main sequence, we demonstrate that models with the same mass and very similar surface properties can have different internal distributions of hydrogen and convective core masses.
We discuss why massive stars expand after the main sequence and the fundamental reasons for why they become red, blue or yellow supergiants.
For the post-main sequence, we demonstrate that small changes in the abundance profile can cause very large effects on the surface properties. We also discuss the effects that produce blue supergiants and the cause of blue loops.
Our models show that massive stars with lower metallicity tend to be more compact due to the combined effect of lower CNO abundances in the burning regions and lower opacity in the envelope.
\end{abstract}


\section{Introduction}

Stars evolve due to changes in their internal chemical composition, ultimately driven by nuclear fusion.
Their interiors are also affected by many other physical processes such as convection, rotation, mass loss and binary interaction \citep[see reviews by ][]{Chiosi1986, Maeder2000, Langer2012}.
Because these processes modify the internal chemical abundance profiles in multiple ways simultaneously, it can be challenging to determine which properties of the stellar interior are primarily driving the overall evolution at any given point in its life.
In this paper we compute stellar models with the aim to (i) devise a qualitative framework to understand the evolution of the surface properties of massive stars and (ii) to isolate the key features of the internal abundance profiles that set the luminosity and effective temperature and drive the evolution of massive stars during different evolutionary stages.

Because our work aims to isolate the key features of the internal abundance profiles that set the luminosity, \ly, and effective temperature, \teff, 
we first review some of the classical results from studies of how physical processes (mass loss, rotation, convection and binary interaction) affect the evolution of \ly and \teff in massive stars.
Stellar evolution models show that mass loss can modify the location of the terminal-age main sequence in the Hertzsprung-Russell (HR) diagram \citep{Chiosi1974, Chiosi1986, Meynet1994, De-Loore1977, Chiosi1978} and favours a lower \teff during core helium burning, near the Hayashi line \citep{Hayashi1962a, Stothers1979, Maeder1987}.
Significant mass loss in higher mass stars can cause evolution to higher \teff during core helium burning if the envelope becomes stripped from the star \citep{Maeder1981b, Sreenivasan1985, Maeder1987, Salasnich1999, Vanbeveren2007, Yoon2010, Georgy2012a, Groh2013, Meynet2015a}.
Stellar rotation is also important for massive star evolution \citep{Heger2000, Maeder2000, Maeder2009}.
In general, moderate rotation favours evolution to higher luminosities and lower \teff during both the hydrogen and helium burning phases \citep{Meynet2000, Chieffi2013}.
Fast rotation can produce evolution towards higher \teff during hydrogen burning if the star becomes chemically homogeneous \citep[e.g.][]{Maeder1987, Langer1992, Yoon2005, Brott2011}.

One of the most important (and uncertain) physical processes that changes the internal abundance profile over time is convective mixing \citep[e.g][]{Shaviv1973, Stothers1973, Maeder1985}.
Both the convective stability criterion and the nature of convective boundary mixing are known to have important impacts on the evolution of \ly and \teff.
For example, a moderate amount of convective core overshooting during core hydrogen burning results in higher \ly and lower \teff at the end of the main sequence and therefore an extended main sequence width in the HR diagram \citep{Maeder1975, Maeder1976, Maeder1981, Alongi1993, Martinet2021}.
The choice of the Ledoux or Schwarzschild criterion for convective stability affects \ly and \teff during core helium burning \citep{Oke1952, Saslaw1965, Stothers1975, Stothers1976, Georgy2014, Georgy2021}.
Convective overshooting in envelopes and the value of the mixing length parameter can also impact the evolution of \teff, particularly for red supergiants \citep{Alongi1991, Chun2018}.
In regions that are stable with respect to the Schwarzschild criterion, but unstable to the Ledoux criterion, semi-convection can occur \citep{Langer1983}. Semi-convection can affect the evolution of \ly and \teff in multiple ways and has been shown to favour core helium ignition as a red supergiant rather than a blue supergiant \citep{Langer1985, Schootemeijer2019}. 

Additionally, interaction with a binary companion greatly complicates the possible evolutionary pathways that stars can follow \citep{Paczynski1967, Sana2012, Moe2017}.
Depending on the nature and outcome of the interaction, binary interaction can have a wide variety of effects on \ly and \teff \citep[see][and citations therein]{Eldridge2017}.
In addition to these evolutionary processes, the initial metallicity is known to play a very important role in the evolution of the surface properties \citep[e.g.][]{Schlesinger1969, Trimble1973, Chiosi1974a, Schaller1992, Langer1995, Groh2019}.
Low metallicity stars are believed to spend more of their post-main sequence evolution in the blue region of the HR diagram, have more extended blue loops \citep{Schaller1992} and produce more blue supergiant supernova progenitors \citep{Langer1991}. 
Other factors that potentially affect the \ly and/or \teff of a star include the presence of strong stellar winds \citep{Langer1989a, Petrovic2006, Grassitelli2021}, magnetic fields \citep{Townsend2005}, turbulent atmospheres \citep{Stothers2003, Cantiello2009, Grassitelli2015} and envelope inflation \citep{Ishii1999, Grafener2012, Sanyal2015, Jiang2015, Grassitelli2018}.

These classical results, combined with more recent studies, have lead to great advances in the understanding of the evolution of massive stars.
However, it is still not always obvious what drives a star to evolve to the blue or the red in the HR diagram, beyond describing what happens in stellar evolution models.
In addition to computing standard stellar evolution models, previous studies have studied this question using a variety of techniques.
The question of why stars evolve to become red giants after core hydrogen depletion has been studied by artificially modifying stellar evolution models \citep{Hoppner1973, Sugimoto2000, Stancliffe2009}.
Other studies have analysed stellar evolution models in terms of the effect of specific features of the internal abundance profile, such as the effect of the hydrogen gradient above the helium core on the surface properties \citep{Schootemeijer2018}.
Another technique involves making artificial changes to a stellar structure model and then using a stellar evolution model to reveal the resulting changes \citep[e.g.][]{Faulkner1966}.
The technique which we will use in this paper, is to combine static stellar structure models with stellar evolution models \citep{Giannone1967, Giannone1967a, Lauterborn1971, Lauterborn1971a, Farrell2020b, Farrell2020a}.
In \citet{Farrell2020a}, we introduced a method called \textsc{snapshot} to compute stellar structure models in hydrostatic and thermal equilibrium and used them to investigate the connections between the core mass, envelope mass and core composition on \ly and \teff. 
We applied these models in several ways, including to study red supergiants \citep{Farrell2020a} and to stars stripped of their envelope through binary interaction \citep{Farrell2020b}.

In this paper, we extend our \textsc{snapshot} method to study the connections between the stellar interior and the surface properties during different evolutionary phases.
Our models allow us to demonstrate the sensitivity of the position of stellar models in the HR diagram in a quantitative way.
These aims are worth pursuing for many important reasons and we list some below.
First, an improved understanding of the cause and effect in stellar evolution models and in the evolution of actual stars can lead to a more satisfactory understanding of how stars evolve.
Second, understanding the key properties that set the luminosity and effective temperature can provide a new way to interpret observations of individual stars and stellar populations in terms of the structural properties that favour a given set of observed properties.
Third, the impact of stars on their immediate surroundings and the rest of the universe through the emission of photons, the release of mechanical energy by stellar winds and the release of chemical elements is regulated by the evolution of the luminosity, \ly, and effective temperature, \teff.
Fourth, the timing of binary interaction and its outcome is affected by the evolution of \ly and \teff.
Additionally, the radius and \teff of massive stars are also relevant for the structure and binding energy of the envelope which has important implications for pulsational pair instability eruptions and other transients.
Fifth, the outputs from stellar evolution models are often counter-intuitive and sensitive to initial conditions.
Finally, this allows us to identify areas for future work to improve stellar evolution models.
These reasons also connect to studies of stellar populations \citep[e.g.][]{Hurley2000, Eldridge2008, de-Mink2013}, the pre-supernova structure and appearance of massive stars \citep[e.g.][]{Heger2003, Groh2013}, supernova explosion properties, massive star nucleosynthesis \citep[e.g.][]{Woosley1995, Chieffi2004, Nomoto2013} and interpreting gravitational wave observations \citep[e.g.][]{Abbott2019}.

In Sec.~\ref{models}, we describe our method and our detailed numerical stellar models.
In Sec.~\ref{theory}, we use current theory of stellar structure and evolution to produce a new framework to understand why stars evolve the way they do. We also present a numerical test that demonstrates clearly why stellar evolution models are so sensitive to small changes in initial conditions.
Following this, we describe the important effects for the two main long-lived nuclear burning stages -- hydrogen burning (Sec.~\ref{ms_results}) and helium burning (Sec.~\ref{he_results}).
We then describe what happens during the short-lived phase between hydrogen and helium burning that often lead to an expansion and the formation of a red (super)giant (Sec.~\ref{sec:hr_gap}).
Finally, we apply our results to explain cause and effect in some representative stellar evolution models (Sec.~\ref{cause_effect_summary}) and discuss other implications of our results (Sec.~\ref{discussion}).

\section{Numerical Stellar Models} \label{models}

\begin{table}
\centering
\caption{Sequences of snapshot stellar structure models (S-) and numerical test models (T-) in the paper.}
\label{table1}
\begin{tabular}{l | l}

\hline
\multicolumn{2}{c}{Main Sequence Models} \\
\hline
Sequence & Isolated property \\
\hline

S-1 & Average mean molecular weight (\muavg) \\ 
S-2 & CNO abundance in the core (CNO$_{\rm core}$)\\ 
S-3 & Fuel supply in the core (Fuel$_{\rm core}$)\\
S-4 & Metal abundance in the envelope (Z$_{\rm env}$) \\
S-5 & Total stellar mass \\
S-6 & Homogeneity of hydrogen profile (H$_{\rm profile}$)\\ 

\hline
\hline
\multicolumn{2}{c}{Core Helium Burning} \\
\hline
Sequence & Isolated property \\
\hline

S-7 &   Distribution of He in the envelope \\ 
S-8 &   Hydrogen gradient in H-shell \\ 
S-9 &   He mass in H-shell with similar gradient \\ 
S-10 &  H abundance in envelope with same H-shell profile \\ 
S-11 &  H abundance in envelope with same gradient \\ 
S-12 &  H abundance in envelope with same inner envelope \\ 
S-13 &  Helium abundance in the core (\ycore) \\ 
S-14 &  Core mass ratio with very shallow hydrogen gradient \\
S-15 &  Core mass ratio with shallow hydrogen gradient \\
S-16 &  Core mass ratio with medium hydrogen gradient \\
S-17 &  Core mass ratio with steep hydrogen gradient \\
S-18 &  Core mass ratio with very steep hydrogen gradient \\
S-19 &  CNO abundance in H-shell (\cnoshell) \\ 
S-20 &  Metal abundance in the envelope ($Z_{\rm envelope}$)\\ 

\hline
\hline
\multicolumn{2}{c}{Expansion after the MS phase} \\
\hline
Sequence & Description \\
\hline

T-21 &   12 \msun stellar evolution model \\ 
T-22 &   Artificially suppressing contraction of the core \\ 
T-23 &   Effect of different \yshell  \\ 
T-24 &   Effect of different \cnoshell \\ 

\hline
\hline
\end{tabular}
\end{table}

In \citet{Farrell2020a} we introduced our method to compute {\sc snapshot} stellar structure models. 
These \textsc{snapshot} models are static stellar structure models in hydrostatic and thermal equilibrium. 
They are a snapshot at just one moment during a star's evolution, so they are not evolving in time. 
The key advantage of \textsc{snapshot} models is that they allow us to systematically isolate the effect of one property of the internal abundance profile at a time, similar to the approach adopted in several previous works \citep{Giannone1967a, Lauterborn1971, Lauterborn1971a, Farrell2020b, Farrell2020a}.
In \citet{Farrell2020a}, we primarily studied the impact of the core and envelope masses on the surface properties of massive stars.
In this work, we expand our method and our focus to a wider range of the key features of internal abundance profiles.
Our \textsc{snapshot} models are computed using the {\sc mesa} software package \citep[r15140, ][]{Paxton2011, Paxton2013, Paxton2015, Paxton2018, Paxton2019}. The steps we take to compute the models are as follows:
\begin{enumerate}
    \item We compute a stellar evolution model at a given mass and metallicity and then save a snapshot at the desired evolutionary stage. The purpose of this is to obtain a starting stellar model which will then be modified. For these evolutionary models, we use a standard set of physical ingredients, the same as described in \citet{Farrell2020b}. However, the exact choices for the physical inputs such as convective overshooting, rotation or even binary interaction are not very important because the models will be modified in the next step. Note that in this study, we don't consider the effects of hydrodynamics, the hydrostatic effect of rotation, magnetic field terms or turbulent pressure.
    \item Once we have an appropriate starting stellar model, we directly modify part of the model file by hand (the .mod file in {\sc mesa}). For example, we might modify the abundance profile in a specific region, or add or remove mass. This can easily be done in a controlled way, varying only one property at a time.
    \item We then insert the model file back into {\sc mesa} to find a solution to the stellar structure equations in hydrostatic and thermal equilibrium with the new abundance profile. As long as the abundance profile or mass was not modified by too much in (ii), we found that \textsc{mesa} usually converged to a stable solution relatively quickly. During this process, we allow convective mixing to briefly take place so that the new solution is consistent with the criterion for convection. Sometimes this causes mixing and changes the chemical profiles inside the model. We also performed tests to verify that our models are in equilibrium by evolving them for a short amount in time to check that they don't change significantly.
\end{enumerate}
Steps (ii) and (iii) are repeated many times as required to construct a series of models in which one feature of the internal abundance profile is changed at a time, e.g. the hydrogen abundance in the envelope (S-10) or the envelope mass (S-16).
The advantage of our method is that we can create numerical experiments indicating how a very specific change at a given point in the internal abundance profile affects the position in the HR diagram. 
This cannot be done by more simplified techniques such as homology relations or polytropic models.
Many different properties of the internal abundance profile can be studied in this way including aspects governed by mixing processes, e.g. the quantity of helium in the hydrogen burning shell, or by mass loss processes, e.g. the mass of the envelope.

\section{A framework to qualitatively understand the evolution of the surface properties of stars} \label{theory}

In Sections \ref{ms_results} -- \ref{discussion}, will use our \textsc{snapshot} modelling approach to investigate the effect of a wide variety features of the internal abundance profile (as listed in Table \ref{table1}).
Before presenting the quantitative results from our numerical models, we find it useful to devise a framework to qualitatively understand the evolution of the radius and luminosity based on the usual equations of stellar structure and, in particular, on energy conservation. 
We describe this framework below.

Classically, expansion and contraction in stars can be encapsulated by \egrav in the equation of energy conservation \citep[e.g.][]{Kippenhahn1990}
\begin{equation} \label{eq_energy}
    \frac{d \lactual}{dm} = \epsilon_{\rm nuc} - \epsilon_{\nu} + \epsilon_{\rm grav} 
\end{equation}
where \egrav can be expressed as
\begin{equation} \label{eq_egrav}
\epsilon_{\rm grav} = - c_{\rm P} T \bigg(\frac{1}{T}\frac{\partial T}{\partial t} - \frac{\nabla_{\rm ad}}{P}\frac{\partial P}{\partial t} \bigg)    
\end{equation}
and where $\epsilon_{\rm nuc}$ is the rate of nuclear energy generation per unit mass, $\epsilon_{\nu}$ represents neutrino losses. 
For clarity, we define \lactual as the actual internal luminosity profile as a function of mass in a star (note that this quantity is often referred to as $L(m)$ in textbooks). 
\egrav expresses the change in the thermodynamic properties of the gas resulting from two possibilities: (i) energy that cannot be removed sufficiently quickly by the energy transport mechanism, in which case the energy remains locked in the gas and (ii) an energy deficiency, in which case the temperature gradient is modified by a change of the stellar structure, causing a contraction.
A local value of $\egrav > 0$ indicates a local contraction, $\egrav < 0$ indicates a local expansion, while $\egrav = 0$ indicates local thermal equilibrium. These expansions or contractions operate on the thermal timescale. Understanding the behaviour of \egrav is critical to understanding why a star evolves to a particular \ly and \teff.

To get an intuitive understanding for what sets \egrav throughout a star as it evolves, we divide the factors that affect the value of \egrav into two components, \lnuc and \lactual. We define \lnuc as the cumulative internal luminosity profile produced by nuclear reactions,
\begin{equation}
    \lnuc(m) = \int_{0}^{m} \epsilon_{\rm nuc} \,dm'.
\end{equation}
\lnuc can be affected by anything that changes the nuclear energy generation rates, i.e. the temperature, density or fuel supply in a nuclear burning region. \lactual can be affected by (i) the hydrostatic structure of the star and (ii) the energy transport within the star. These are described by the equations for hydrostatic equilibrium and energy transport respectively \citep[e.g.][]{Kippenhahn1990},
\begin{equation}
    \frac{dP}{dm} = - \frac{G m}{4 \pi r^2}
\end{equation}

\begin{equation} \label{eq_et}
    \frac{dT}{dm} = - \frac{G m T}{4 \pi r^4 P} \nabla
\end{equation}

\begin{equation}
\nabla_{\rm rad} =  \frac{3}{16 \pi a c G}\frac{\kappa \lactual P}{m T^4}
\end{equation}
where $\kappa$ is the opacity and all other variables have their usual meaning. For transport by radiation, $\nabla = \nabla_{\rm rad}$ in Equation \ref{eq_et}, while for transport by convection $\nabla$ is equal to $\nabla_{\rm ad}$ or the gradient that is obtained from mixing-length theory. \lactual can be affected by any property that affects hydrostatic equilibrium and energy transport including the mass, the opacity or the presence of convection. 
Rearranging Equation \ref{eq_energy}, and assuming $\epsilon_{\nu} << \epsilon_{\rm nuc}$ (which is valid for the vast majority of a star's lifetime), we get
\begin{equation} \label{eq_rearr}
    \epsilon_{\rm grav} = \frac{d \lactual}{dm} - \epsilon_{\rm nuc} = \frac{d \lactual}{dm} - \frac{d \lnuc}{dm}
\end{equation}
Starting in thermal equilibrium, $\egrav = 0$, an increase in $d \lnuc/dm$ or decrease in $d\lactual/dm$ will lead to $\egrav < 0$ and expansion, and vice versa. 
The picture can be simplified by considering that an increase/decrease in \lnuc or \lactual at a given point in the star will also result in a local increase/decrease in $d \lnuc/dm$ or $d\lactual/dm$ respectively. 

When a star is in perfect hydrostatic and thermal equilibrium, $d\lnuc/dm = d\lactual/dm$ and $\lactual = \lnuc$ at all points in the star. 
Any change that causes either an increase of \lnuc or a decrease of \lactual will favour evolution to a larger radius (usually lower \teff).
Conversely, any property that causes a decrease of \lnuc or an increase of \lactual will favour evolution to a smaller radius (usually higher \teff). 
The evolution of the luminosity is determined by how the surface value of \lactual changes when the star relaxes to thermal equilibrium. 
It typically increases with changes that increase \lactual, and vice versa. 
However, the change in luminosity can be difficult to predict a priori due to the possibility of the formation of convective zones. 
In summary, stars
\footnote{Note that we are not considering stars that are out of hydrostatic equilibrium, e.g. pulsating stars or luminous blue variables in eruption, which behave very differently to stars evolving on nuclear or thermal timescales.}
can contract or expand on nuclear or thermal timescales due to changes in e.g.
\begin{enumerate}
	\item The temperature, density or fuel supply of a nuclear burning region (affects \lnuc)
	\item The hydrostatic structure of the star e.g. a decrease in the envelope mass due to mass loss (affects \lactual)
	\item The efficiency of energy transport e.g. a change in opacity or the presence of convection (affects \lactual)
\end{enumerate}
Any change to the internal abundance profile, the hydrostatic structure or the energy transport that causes a mismatch between \lnuc and \lactual will change the surface properties of a star.
The star will relax towards thermal equilibrium and, therefore, a new \ly and \teff.

\begin{figure*} \centering
\includegraphics[width=\linewidth]{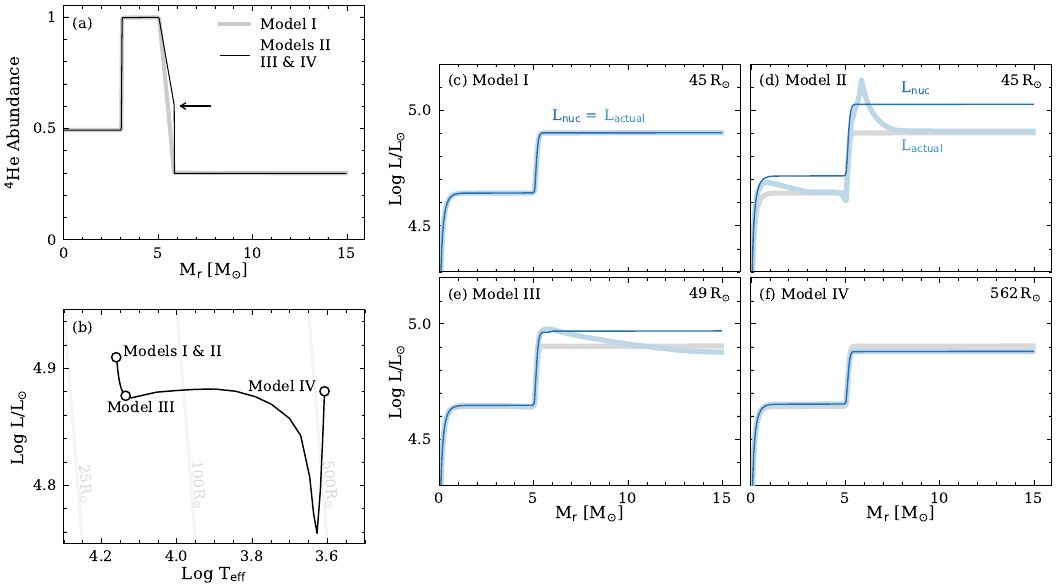}
\caption{A numerical test to examine the response of a stellar model to changes in its abundance profile. We plot the internal helium abundance profile and the surface properties for the following four models. \textit{Model I}: the original blue supergiant in thermal equilibrium and in which helium will be inserted. \textit{Model II}: stellar model immediately after the helium is inserted to the H-shell by hand, indicated by the black arrow in (a). \textit{Model III}: a few timesteps after Model II after the model has expanded to a radius of $49~\rsun$. \textit{Model IV}: when the model has reached thermal equilibrium as a red supergiant. For each model, we plot the cumulative nuclear energy generation profile \lnuc and the luminosity imposed by hydrostatic equilibrium \lactual in panels.}
\label{fig:test}
\end{figure*}

\begin{figure*} \centering
\includegraphics[width=\linewidth]{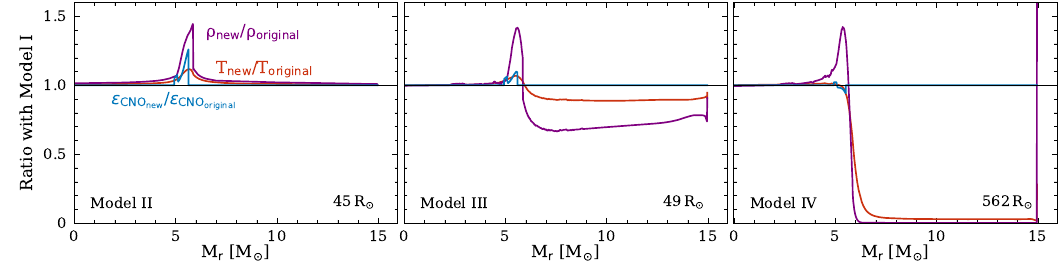}
\caption{For Models II, III and IV from the test plotted in Fig. \ref{fig:test}, we plot the ratio of the internal profiles of temperature, density and $\epsilon_{\rm nuc}$ from CNO burning (scaled by a factor of 0.15) with Model I.}
\label{fig:test_interiors}
\end{figure*}

\begin{figure} \centering
\includegraphics[width=\linewidth]{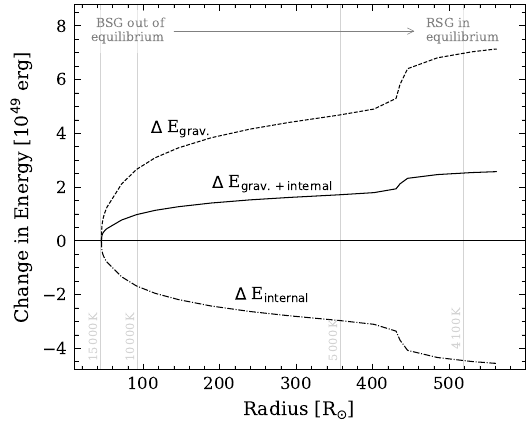}
\caption{The change in the total internal, gravitational and total (internal + gravitational) energy of the star from the test in Fig. \ref{fig:test} as it expands from a blue to a red supergiant and eventually reaches thermal equilibrium on the right-hand side of the figure. Quantities are defined in the usual way e.g. \citet{Kippenhahn1990}. The bump at $\sim 440~\rsun$ is due to convection in the envelope.}
\label{fig:test_energies}
\end{figure}

We now describe a numerical test to demonstrate how a star responds when $\lnuc~\neq~\lactual$ and why stellar evolution models are sometimes so sensitive to small changes.

\begin{enumerate}
    \item The initial stellar structure model corresponds to  $15\,\msun$ blue supergiant at the middle of the core helium burning phase (Fig.~\ref{fig:test}ab). 
    This model is in both hydrostatic and thermal equilibrium and $\lnuc~=~\lactual$ at all points in the star (Fig.~\ref{fig:test}c). 
    \item We perturb the initial model by modifying the internal abundance profile by hand in the region of the hydrogen burning shell, replacing a small amount of hydrogen with helium (Fig.~\ref{fig:test}a). 
    This new \new{abundance profile} could, for instance, correspond to what one would obtain assuming a slightly different implementation of the (uncertain) internal mixing processes. 
    We then put the \new{modified model file} into \mesa to find a solution to the stellar structure equations, enforcing hydrostatic equilibrium, and study how it relaxes to thermal equilibrium.
    \item  We obtain Model II, which is initially out of thermal equilibrium and $\lnuc~\neq~\lactual$ (Fig.~\ref{fig:test}d).
    The surface properties of Model II have not changed compared to Model I as the changes in the interior have not yet propagated to the surface (Fig.~\ref{fig:test}b). 
    The larger helium abundance of Model II increases the mean molecular weight (\muu) in the H-shell compared to Model I.
    \new{For a higher \muu, a higher temperature and/or density is required to maintain hydrostatic equilibrium due to the equation of state.}
    In this case, both the temperature and density in the H-shell are higher in Model II compared to Model I (Fig.~\ref{fig:test_interiors}a). 
    This increases the rate of nuclear energy generation in the H-shell, therefore increasing \lnuc. 
    Thus the value of \lnuc in the envelope is larger in Model II than Model I (Fig.~\ref{fig:test}d). 
    \item The value of \lnuc is greater than \lactual above the H-shell (Fig.~\ref{fig:test}d). 
    This means the luminosity entering the envelope 
    is greater than the luminosity that can be transported by the envelope with its current structure. 
    The excess energy that cannot be transported will be absorbed into the envelope and cause it to cool and expand. 
    As long as $\lnuc~\neq~\lactual$, the star will continue to expand and move to the right in the HR diagram (Fig. \ref{fig:test}b). 
    The expansion will stop when the energy produced in the H-shell is exactly equal to the energy that the envelope can transport, i.e. $\lnuc~=~\lactual$, and the envelope has a new structure.
    \item As the star expands, the values of \lnuc and \lactual change. \lnuc gradually decreases because the expansion of the envelope causes the temperature and density in the outer regions of the H-shell to decrease (Fig.~\ref{fig:test_interiors}). 
    In the envelope, \lactual initially decreases and then subsequently increases. 
    This is reflected in the change of the surface luminosity between Model III and IV (Fig.~\ref{fig:test}b). 
    The initial decrease is a result of the decrease in the temperature gradient dT/dr which decreases the efficiency of energy transport by radiation and, hence, decreases \lactual. 
    The decrease of \lactual actually makes it more difficult for the star to reach thermal equilibrium because it causes the difference between \lnuc and \lactual to increase. 
    As the radius increases, an increasing proportion of the envelope begins to transport energy by convection. 
    This is a result of cooling of the envelope and an increase in $\nabla_{\rm rad}$. 
    As convection is more efficient \new{than radiation} at transporting energy, it favours an increase in \lactual (and \ly), which occurs once enough of the envelope becomes convective.
    It is important to note that the formation of the convective envelope is an effect of the expansion, not the cause. 
    The star will expand regardless of the onset of convection.
    \item Another way to think about the expansion is in terms of the change in the internal and gravitational energy (Fig.~\ref{fig:test_energies}). 
    As discussed above, in Model II the H-shell is producing more luminosity than the envelope can transport. 
    The resulting cooling and expanding of the envelope corresponds to a decrease in the total internal energy of the star of $4.5 \times 10^{49}$erg and an increase in its total gravitational potential energy of $7.0 \times 10^{49}$erg, reflecting the virial theorem for stars. 
    As a result, the total (gravitational + internal) energy of the star increases by $2.5 \times 10^{49}$erg. 
    This net increase comes from nuclear energy generation in the H-shell, so that energy is conserved in total. 
    The increase in the total energy is almost entirely due to the increase in the energy of the envelope. 
    In general, processes which increase the total energy of the envelope also tend to favour an increase in the total stellar radius.
\end{enumerate}
In the following sections, we will use \lnuc and \lactual to explain what sets the values of $L$ and \teff during core hydrogen burning (Sec.~\ref{ms_results}), core helium burning (Sec.~\ref{he_results}) and the expansion across the HR diagram between between the hydrogen and helium burning phases (Sec.~\ref{sec:hr_gap}).

\section{The Core hydrogen burning phase} \label{ms_results}

The Main Sequence phase is the most studied and well known phases of the evolution of stars in general \citep[e.g.][]{Schwarzschild1958a, Kippenhahn1990}.
In this section, we want to (i) verify that our \textsc{snapshot} models can recover well known results for stars on the MS from simpler methods e.g. homology relations, and to (ii) go beyond what can be done with these simpler approaches by presenting some numerical experiments that shed new light on otherwise well-known behaviour of stars during the MS.

\subsection{What sets the luminosity and effective temperature of a star on the main-sequence?}

\begin{figure*} \centering
\includegraphics[width=1.0\linewidth]{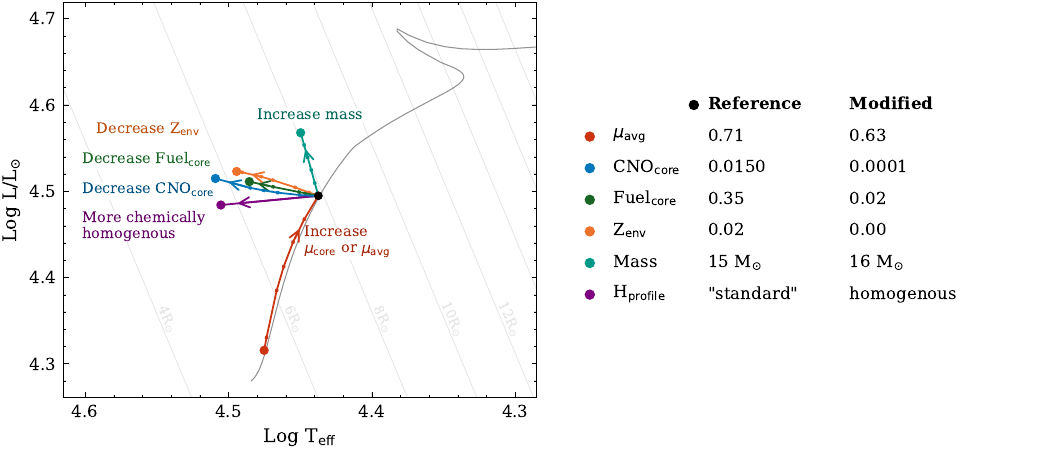}
\caption{The coloured lines represent sequences of models (S-1 to S-6) in which the impact of different properties on the surface properties of a 15 \msun main sequence star are isolated. These are the average mean molecular weight (\muavg), the CNO abundance in the convective core (CNO$_{\rm core}$), the fuel supply in the core (Fuel$_{\rm core}$), the metal abundance in the envelope (Z$_{\rm env}$), the total stellar mass and the homogeneity of the hydrogen abundance profile (H$_{\rm profile}$). An evolutionary track of a 15 \msun star with solar metallicity (Z = 0.020) is plotted in grey for reference. See the online supplementary material for plots of the internal abundance profiles.}
\label{fig:ms_hrd}
\end{figure*}

\begin{figure} \centering
\includegraphics[width=1.0\linewidth]{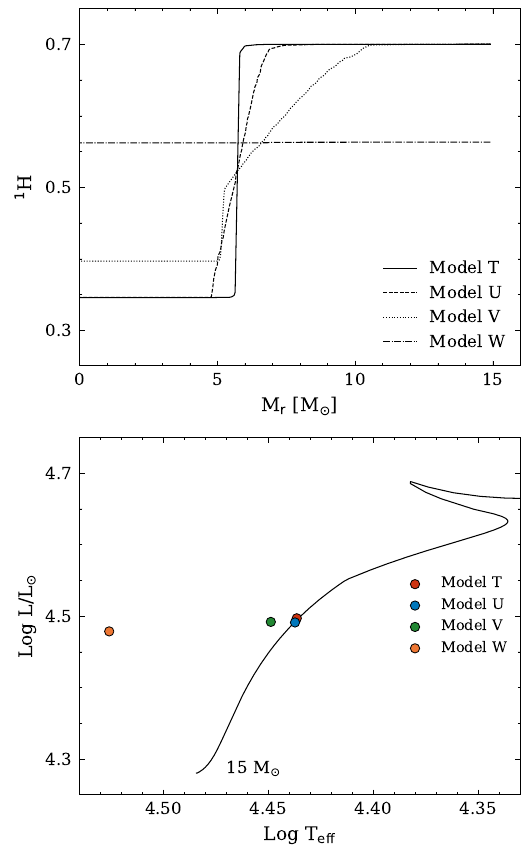}
\caption{\textit{Upper panel:} Hydrogen abundance profiles of four 15 \msun stellar models with the same total hydrogen mass (i.e. same \muavg), but with different internal distributions. \textit{Lower panel:} Location in the HR diagram of the four models from the upper panel with an evolutionary track of a 15 \msun plotted for reference.}
\label{fig:ms_degeneracy}
\end{figure}

We \new{investigate} six key properties that set \ly and \teff of massive main sequence stars. These are the average mean molecular weight (\muavg), the CNO abundance in the convective core (CNO$_{\rm core}$), the fuel supply in the convective core (Fuel$_{\rm core}$), the metal abundance in the envelope (Z$_{\rm env}$), the total stellar mass and the homogeneity of the hydrogen abundance profile (H$_{\rm profile}$).
To demonstrate the effect of each of these properties, we model their impact on the surface properties with reference to a model of a 15~\msun solar metallicity star at the middle of the main sequence phase.
To do this, we compute a sequence of models in which only one property of the stellar interior is changed at a time, with the exception of the H$_{\rm profile}$ in which both the hydrogen and helium abundance profiles change.
The mean molecular weight and the fuel supply are unusual because they are not simple features of the internal abundance profile.
To isolate these properties, we introduce an artificial element called `non-burning hydrogen' which has the mean molecular weight and opacity properties of hydrogen but doesn't participate in nuclear burning.
We achieve this in \textsc{mesa} by modifying the list of isotopes considered in the model to include an artificial isotope with the chemical properties of hydrogen. 
However, as we do not add this artificial isotope to any of the nuclear reaction networks, it doesn't participate in nuclear reactions.
We verified that this isotope behaves as expected in a stellar evolution model.

Fig. \ref{fig:ms_hrd} presents the six sequences of models in the HR diagram as well as an evolutionary track of a 15 \msun star at Z = 0.020 for comparison.
The points on each line indicate the location of each individual snapshot model and the arrows indicate the direction of increase/decrease.
The legend on the right lists the numerical values of each property for the reference model plotted in black and for the modified models indicated by the large coloured points at the other end of each model sequence. 
See the online supplementary material for the internal abundance profiles of the model sequences in Fig. \ref{fig:ms_hrd}.
In the following paragraphs, we discuss the effects of each of these properties in detail and consider how they affect \lnuc and \lactual.

To change the average mean molecular weight (\muavg), we convert some fraction of the helium in the convective core to `non-burning hydrogen'.
This has the desired effect of reducing the mean molecular weight without modifying the fuel supply.
It does have the small side effect of changing the opacity in the interior but this has a quantitatively small effect on the overall structure of the stellar model.
Note also that while the set of models plotted in Fig. \ref{fig:ms_hrd} does only modify $\mu$ in the convective core, we also computed models in which $\mu$ is changed both in and above the core and found similar results.
Through the effect of the equation of state, a higher \muavg favours a larger radius $R$, lower \teff and higher luminosity \ly.
This result can be understood by considering how \muavg affects \lnuc.
For a higher $\mu$, the equation of state requires a higher temperature and/or density to maintain a given pressure gradient and, therefore, to maintain hydrostatic equilibrium.
Due to the higher temperature and density in the nuclear burning region, the rate of nuclear energy generation increases, thus increasing \lnuc.
This increase in \lnuc favours a larger radius, as discussed in Sec.~\ref{theory}.
The increase in the surface luminosity \ly with \muavg is due to the higher temperature and density in the central burning region once the model has relaxed to thermal equilibrium. 
One can derive a consistent conclusion using homologous relations, where $L \propto \mu^{4}$ \citep[e.g.][]{Kippenhahn1990}. Indeed, one has $L(\mu = 0.71)/L(\mu = 0.63) \approx (0.71/0.63)^4$.

We modify the CNO abundances in the core by scaling down the abundances of all the C, N and O isotopes from the reference model (with a solar metallicity) and replacing them with hydrogen.
The primary impact of this on the stellar structure is to modify the abundance of CNO isotopes available to act as catalysts in the CNO cycle.
We find that a lower CNO abundance in the core (CNO$_{\rm core}$) favours a smaller $R$, higher \teff and higher \ly.
The effect on the radius can be understood by considering that a decrease in CNO$_{\rm core}$ decreases \lnuc (because the nuclear energy generation rates scale with the abundance of CNO elements, $\epsilon_{\rm nuc} \propto X_{\rm CNO}$) which favours a smaller radius.
So, why does the luminosity increase?
One might have guessed that the higher temperatures in the core required to produce enough luminosity to support the star would exactly balance the effect of the lower CNO abundances.
Our models show that the hydrostatic structure of a star in thermal equilibrium with a lower abundance of CNO elements in the core has an internal temperature and density profile that drops off slower as you move away from the center of the star.
This results in a higher energy generation rate in the outer parts of the nuclear burning region.
An interesting consequence of this is that if you remove fuel, either hydrogen or CNO abundances, from the core, and keep other quantities constant, a star will actually become more luminous.

The fuel supply in the core (Fuel$_{\rm core}$) can be modified by converting some fraction of the hydrogen in the core to `non-burning hydrogen'. 
We find that a decrease in Fuel$_{\rm core}$, holding everything else constant, favours a smaller $R$, higher \teff and higher \ly.
Due to the effect on nuclear energy generation a decrease of Fuel$_{\rm core}$ will decrease \lnuc, favouring a smaller radius.
As in the case of the CNO$_{\rm core}$, because a lower value of Fuel$_{\rm core}$ favours a more compact star in thermal equilibrium, the temperatures and densities throughout the nuclear burning region (in particular the outer parts) are slightly higher.
This results in a higher surface luminosity.

A lower abundance of metals in the envelope (Z$_{\rm env}$) favours a smaller $R$, higher \teff and higher \ly.
This is due to the effect of Z$_{\rm env}$ on the opacity of the envelope.
A decrease in opacity affects the energy transport and increases \lactual in the envelope, favouring a smaller radius, higher \teff and higher luminosity.

We isolate the effect of the mass by keeping the same abundances at a given normalised mass fraction but modifying the total stellar mass.
A larger mass causes a larger $R$, higher \teff and higher \ly.
While this is very well-known (especially at the zero-age main sequence) and relatively intuitive to understand, it is also worth understanding the result in terms of \lnuc and \lactual, especially to contrast to helium burning stars later on.
A larger mass requires a higher luminosity to maintain hydrostatic equilibrium, resulting in an increase in \lactual. 
This actually favours a contraction to a smaller radius. 
However due to the magnitude of the increase of \lactual, the contraction in the center of the star is large enough to significantly increase the core temperature and the nuclear energy generation rate, causing a feedback effect on \lnuc.
The balance between the competing effects on \lnuc and \lactual results in an overall increase in the radius with mass.
We will come back to this point when discussing the effects of increasing the envelope mass of a core helium burning star. 

Finally, we compare the reference model to a model with the same total mass of hydrogen and helium, but distributed homogeneously throughout the star. While this does increase the fuel supply in the core, it is still a good representation of the effect of the distribution of hydrogen throughout the star (H$_{\rm profile}$). 
The abundance profile of hydrogen in the envelope affects the opacity.
Similar to a decrease in Z$_{\rm env}$, the decrease in opacity allows a higher luminosity to be transported and increases \lactual in the envelope, favouring a smaller radius and higher \teff.
Our models indicate that the hydrogen profile has very little impact on the luminosity.
We elaborate further on the impact of the internal distribution of hydrogen in Sec. \ref{ms_results_deg}.

\subsection{Connection to overshooting and metallicity effects}
Given the above discussion, we can understand why models with moderately high overshooting or rotation evolve to lower \teff at the terminal-age main sequence. When these models reach the lowest value of \teff on the MS, internal mixing has allowed a larger proportion of the hydrogen in the star to be converted to helium. As a result, the value of \muavg is larger. As we have discussed above, a larger value of \muavg favours a larger radius and lower \teff. Another way of looking at this is to imagine picking up a model with higher overshooting at the TAMS and compressing it so that it is located at the TAMS location of a model with lower overshooting (at higher \teff). The star would then be out of thermal equilibrium as \lnuc would be larger than \lactual. This would favour an expansion back to larger radii and lower \teff. For more fully mixed models, corresponding to strong rotation or very high overshooting, the competing effect of opacity due to the amount of hydrogen near the surface of the star dominate over the effect of \muavg, favouring a smaller radius and higher \teff. It's also interesting to note that the isolated effect of metallicity on stellar structure of stars at this mass is composed of two significant components, the CNO abundance in the core and the metal abundance in the envelope. Our models show that the effect of opacity is not the only reason that lower metallicity stars have a higher \teff.

\subsection{Degeneracy between internal hydrogen profile and surface properties} \label{ms_results_deg}

We now examine in more detail the effect of distribution of hydrogen/helium within the star, holding the total mass of hydrogen/helium constant. Fig.~\ref{fig:ms_degeneracy} compares four core hydrogen burning stellar models with the same mass (15 \msun) and the same total helium mass, i.e. same \muavg, but with different internal distributions of hydrogen. All of the models are in hydrostatic and thermal equilibrium. Models T, U and V each have different internal distributions of hydrogen (and helium) and different convective core masses. However they have very similar surface properties, differing by at most 0.01 dex in \logteff and 0.002 in \logllsun. These models demonstrates a degeneracy between the internal distribution of hydrogen and the surface properties for stars with the same total mass and \muavg. The degeneracy applies only in the limit that the effects of opacity in the outer layers of the envelope do not dominate. When the star is more fully mixed and homogeneous (Model W), the effects of opacity in the outer layers of the star dominate and it has a significantly higher \teff. 
Our results here are consistent with conclusions from simplified models of main sequence stars \citep[e.g][]{Schwarzschild1958}, but we feel that it is worth re-emphasising and demonstrating this point with more detailed models.

This degeneracy suggests that a given \ly and \teff could correspond to a lower mass star with higher \muavg or a higher mass star with a lower \muavg. 
Based on the surface properties alone, it may only be possible to constrain a star to a range of allowed total masses and total hydrogen/helium masses. 
The degeneracy could be broken by asteroseismology which would provide a better understanding of internal mixing processes.
It also suggests that current and future asteroseismology studies with larger sample sizes and higher mass stars will be useful to improve our understanding of convective boundary mixing and rotational mixing beyond what is possible using \ly and \teff alone.
Further study is required to properly quantify this degeneracy across different masses and evolutionary states.

\section{The Core Helium burning Phase} \label{he_results}

The evolution of massive stars after the main sequence is complicated.
Some of the fundamentals are well understood after decades of research, however many important questions remain unanswered.
Stellar evolution models are one of the key tools used to tackle these questions.
Understanding exactly what drives a post-main sequence massive stellar model to a blue or red supergiant solution is often very difficult. 
Stars can evolve to the right and left in the HR diagram in ways that are difficult to understand, sometimes executing loops. 
The stellar interior changes in multiple ways simultaneously, which makes it difficult to distinguish cause and effect.
In addition, very small changes to the stellar interior from earlier evolutionary phases can have a significant effect on the subsequent evolution \citep{Iben1974, Weiss1989, Chin1990, Maeder1994, Ritossa1996}. 
For this reason, Rudolf Kippenhahn referred to the post main sequence stage as a ``sort of magnifying glass, also revealing relentlessly the faults of calculations of earlier phases'' \citep{Kippenhahn1990}. 
In this section, we isolate and analyse some of the important features of the internal abundance profile that set \ly and \teff of massive stars during core helium burning. 
Because stellar interiors can get quite complicated and are ultimately described by a set of vectors describing the internal abundance profiles of each isotope, we cannot fully describe every feature of the internal abundance profile that affects the surface properties.
Instead, we try to select the important properties that are modified by evolutionary processes.
These are the effect of the helium abundance in the hydrogen shell, the hydrogen abundance in the envelope, the helium abundance in the core, the core mass ratio, the CNO abundance in the hydrogen shell and the metallicity in the envelope which we discuss in Sections \ref{sec:yshell} - \ref{sec:cnoshell} below.

\subsection{Helium Abundance Profile in the Hydrogen Burning Shell} \label{sec:yshell}
The fact that models with different hydrogen profiles at the interface between the core and the envelope impacts \teff has been known since some of the earliest stellar models were computed \citep[e.g.][]{Lauterborn1971a}. 
Models with steeper hydrogen gradients are found to favour a bluer star while a shallower gradient favours a redder star \citep{Stothers1968, Robertson1971, Fricke1971, Lauterborn1971, Stothers1976, Schlesinger1977, Langer1985, Walmswell2015, Schootemeijer2018, Schootemeijer2019}.
The reason why for this effect is not immediately obvious.
An investigation by \citet{Walmswell2015} isolated the effects of the hydrogen profile on the opacity, mean molecular weight and fuel supply and how each affected \teff.
They found that the increased mean molecular weight and decreased effect of opacity associated with a shallower hydrogen gradient favour a redder star, the reduced fuel supply favours a bluer star, and concluded that the effect of the increased mean molecular weight dominates.
Following this, one may still wonder why a gradient of mean molecular weight at the core/envelope interface affects \teff like this.
In this section, we investigate this question.

To encapsulate the abundance profile of hydrogen at the core/envelope interface, previous stellar evolution studies have used the term ``hydrogen gradient''.
This is a simple quantity that is useful for some purposes.
However, as is well known, the internal abundance profiles can be quite complicated e.g. due to convective shells during the main sequence phase, semi-convection \citep{Langer1985}, mass gainers \citep{Braun1995} and mergers \citep{Glebbeek2013}.
Therefore, we will instead refer to this in a more general way as the helium abundance profile in the region of the hydrogen burning shell, denoted by \yshell.
We will demonstrate why we choose to do this below.

In stellar evolution models, \yshell is primarily set by the receding convective core during the MS phase \citep[e.g.][]{Robertson1971} and by internal mixing between the end of core hydrogen burning and the beginning of core helium burning \citep[e.g.][]{Langer1985}. 
During core helium burning, \yshell evolves as the hydrogen shell burns through the layers at the base of the envelope. 
Several different physical processes can modify the shape of the profile.
Firstly, the implementation of convective core overshooting during the main sequence phase and choice of the free overshooting parameter (\aov or \fov) can cause a steeper or shallower gradient of hydrogen and helium in the H-shell \citep{Stothers1991b, Langer1991}.
Semi-convective mixing between core hydrogen depletion and the beginning of helium burning can also have an important effect \citep{Langer1985, Langer1991, Schootemeijer2019}.
A combination of reasonable choices for the free parameters for convective overshooting and semi-convective efficiency can produce a very wide variety of hydrogen/helium profiles in the H-shell \citep{Schootemeijer2019}.
Rotation has also been shown to modify the helium profile \citep{Maeder2001} and this depends on the initial rotational velocity and choice of the diffusive mixing parameter.
In binary systems, mass gainers and stellar mergers during the MS and the post-MS \citep[e.g.][]{Braun1995, de-Mink2013, Glebbeek2013} and the reorganisation of the star after a merger event could produce abundance profiles that differ significantly to single stars.
In addition, current 1D stellar evolution models likely do not fully or accurately capture the mixing processes that affect \yshell. These instabilities are inherently multi-dimensional as shown by \citet{Cristini2017, Horst2021}.
All of this is to say that the helium abundance profiles in the hydrogen shell that are produced in stellar evolution models are subject to many different uncertain processes.

\begin{figure*} \centering 
\includegraphics[width=\linewidth]{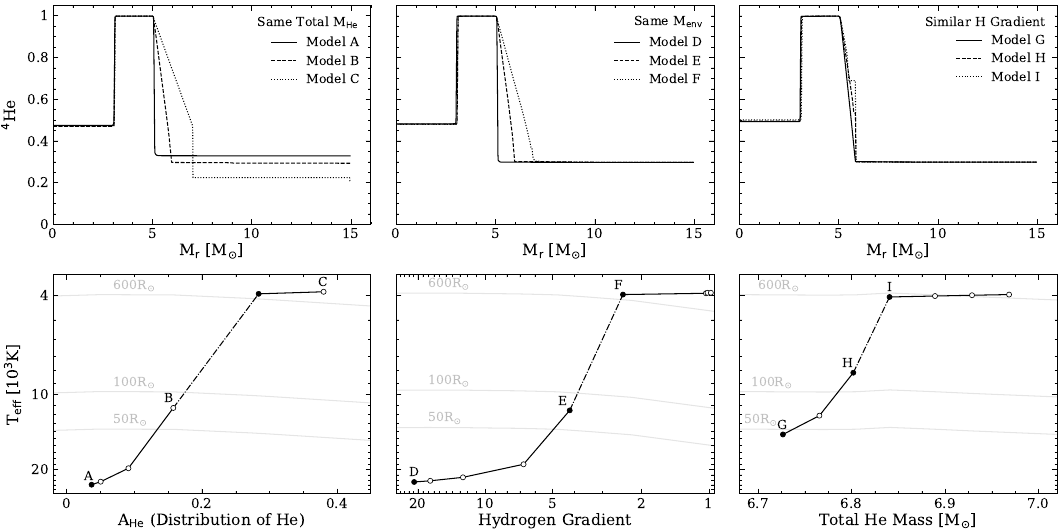}
\caption{Effect of the helium abundance profile in the shell on 
\ly and \teff of core helium burning stars. The upper panels show the internal helium abundance profiles for three snapshot models. The lower panels show the value of \teff for the models from the upper panel (black circles), as well as several similar intermediate models that are excluded from the upper panel for clarity (white circles). The dash-dot lines indicate the bi-stability transition between a blue and red supergiant.
\textit{Left Column}: Models with the same total mass of helium mass, distributed differently throughout the envelope (S-7). 
\textit{Middle Column}: Models with the same envelope mass and with different gradients of hydrogen and helium in the hydrogen burning shell (S-8). 
\textit{Right Column}: Models with similar hydrogen gradients and different masses of helium in the H-shell (S-9).}
\label{fig:yshellteff}
\end{figure*}

We compute several sets of snapshot stellar structure models that isolate the effect of \yshell on \ly and \teff, presented in Fig. \ref{fig:yshellteff}. 
Each set of models isolates \yshell in different ways, keeping some other quantity constant. 
\new{Due to the on-off nature of convection in our stellar models, there is a sudden transition between a blue and a red supergiant solution (connected dash-dot lines in Fig.~\ref{fig:yshellteff}), similar to the one found in \citep{Farrell2020b}.}
The first set of models in Fig. \ref{fig:yshellteff}, Models A, B and C have the same total mass of helium but distributed differently throughout the envelope.
We parameterise the distribution of helium in the envelope in terms of a new parameter
\footnote{\new{$A_{\rm He}$ is defined in terms of two imaginary distributions of helium in the envelope, $A_1(m) = M_{\rm He, env}/\menv$ (a constant value), and $A_2(m)$ which is a step profile with a value of $0$ if $M_r > \mcore + M_{\rm He, env}$, else 1. Then $A_{\rm He} = \int |A_1(m) - Y(m)| dm / \int |A_1(m) - A_2(m)| dm $, where quantities are integrated over the envelope (where $X > 10^{-4}$) and $Y(m)$ is the actual helium abundance profile.}} 
$A_{\rm He}$, where a value of $A_{\rm He} = 0$ means helium is distributed perfectly evenly throughout the envelope and $A_{\rm He} = 1$ means all of the helium is concentrated at the base of the envelope and all the hydrogen is at the surface (quite an artificial situation).
Models with a higher proportion of helium distributed towards the base of the envelope favour redder stars with larger radii, while those with more evenly distributed helium favour bluer stars with smaller radii.
For different distributions of helium, the effective temperature can range from 23~000~K to 4~000~K.

The second set of models, Models D, E and F, have the same core mass (defined here as where $X < 10^{-4}$) and envelope mass and isolate the effect of a linear abundance profile of hydrogen/helium. We choose to parameterise each model in terms of the hydrogen gradient as defined by \citet{Schootemeijer2018}. Models D, E and F indicate clearly that a shallower hydrogen gradient favours a redder star with a larger radius, while a steeper gradient favours a bluer star with a smaller radius. This reproduces the well-known results regarding the hydrogen gradient that we discussed above. For moderate changes in the gradient, our models indicate that the radius can change by a factor of $\sim~20$ and \teff can vary from $22\,000\,$K to $4\,000\,$K.

The third set of models, Models G, H and I, have similar hydrogen gradients consisting of the same drop in hydrogen abundance between the core and the envelope over the same mass interval, but different total masses of helium.
Despite the fact that the hydrogen gradient is very similar, the value of \teff and the stellar radius can vary quite significantly for models with a different mass of helium in the shell.
Although this phenomenon is well known from stellar evolution models, Models H and I clearly demonstrate that very small changes in the stellar interior make a huge difference in \teff.
Although they differ by just $0.05~\msun$ in their total mass of helium, a tiny fraction of the total stellar mass of $15~\msun$, but E is a blue supergiant and F is a red supergiant.
The sharp transition between blue and red supergiants due to a small change in stellar structure is similar to the one discussed in \citet{Farrell2020a}.
The difference between Models G, H and I is smaller than the typical uncertainties in the internal mixing in a stellar evolution model and yet they differ in radius by a factor of 12.
As they are located in different parts of the HR diagram, they may be interpreted as having different evolutionary histories, despite their very similar internal structure.
Although it has been emphasised by many others, we would like to reiterate the importance of being cautious when using the results of an individual stellar evolution model.
A related consequence of these results is that the effect of the helium abundance profile in the shell on the surface properties cannot easily be represented by a single parameter.
Even for a very similar hydrogen gradient \citep[e.g. as defined by][]{Schootemeijer2019}, \teff can vary significantly due to the effect of the helium profile (compare our models H and \& I).

To understand why the helium abundance profile has these effects on \teff, consider how \yshell affects \lnuc and \lactual.
An increase of the helium abundance in the region of the H-burning shell causes an increase in the mean molecular weight and, through the equation of state, in the temperature and density (the same effect demonstrated in detail in the numerical test in Sec. \ref{theory}).
This causes an increase in the nuclear energy generation rate in the hydrogen burning shell, increasing \lnuc.
As discussed in Sec. \ref{theory}, an increase in \lnuc favours a larger radius and, in this case, a redder star.
Any process that causes an increase in the helium abundance in the nuclear burning region of the hydrogen burning shell favours a redder star with a larger radius
\footnote{We can also consider the framework of \lnuc and \lactual in the context of the effects of the fuel supply and the opacity on \teff found by \citet{Walmswell2015} (although note that the effect of the mean molecular weight dominates the impact on the overall surface properties).
A decreased fuel supply causes a decrease in \lnuc which favours a bluer star.
A decreased opacity in the H-burning shell initially favours a local increase in \lactual, causing a local contraction in the hydrogen burning shell which actually increases \lnuc.
The increase in \lnuc dominates, favouring a redder star.
We found the same qualitative results for the effect of the mean molecular weight and fuel supply on the \teff of core hydrogen burning stars in Sec. \ref{ms_results} as \citet{Walmswell2015} did for core helium burning stars.}.

\subsection{Hydrogen Abundance in the Envelope} \label{sec:xenv}
The abundance profile of hydrogen in the envelope of a core helium burning star (\xenv) can be affected by several different factors.
The initial abundance of hydrogen and helium can vary with metallicity.
Internal convective zones and semi-convective mixing in the envelope during the transition between core hydrogen and core helium burning can also alter the hydrogen profile in the envelope.
This can also happen, for instance, if a star becomes a red supergiant and its convective envelope extends deep enough to a region of decreased hydrogen abundance in a dredge-up episode.
Accretion of material from a binary companion with a different surface abundance or a merger event could also modify the envelope abundance.

\begin{figure*} \centering
\centering
\includegraphics[width=\linewidth]{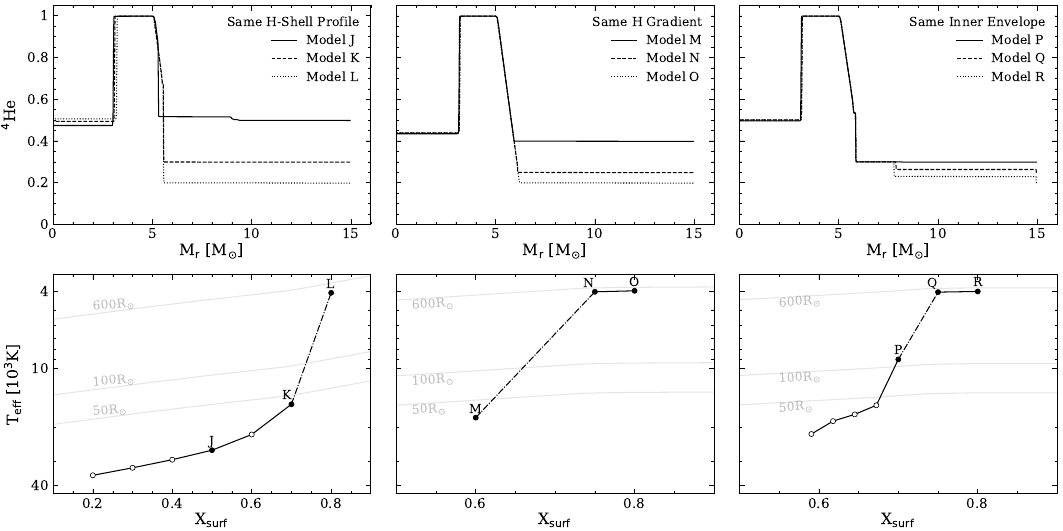}
\caption{Effect of the Hydrogen profile in the envelope on \teff for helium burning stars. \textit{Left Column}: Models with the same hydrogen profile and different abundances in the envelope (S-10). \textit{Middle Column}: Models with the same hydrogen gradient and with different surface abundances (S-11). \textit{Right Column}: Models with different helium abundances in the outer envelope (S-12).}
\label{fig:xenvteff}
\end{figure*}

As in Sec.~\ref{sec:yshell}, we compute sets of snapshot stellar structure models to isolate the effect of \xenv on \teff for models with the same \mcore, \menv, \yc and abundance profile in the H-shell, presented in Fig. \ref{fig:xenvteff}.
For clarity, we choose to plot the helium abundance profiles to demonstrate that the helium cores are the same.
Models J, K and L have the same (or very similar) abundance profiles in the H-shell but different profiles in the envelope.
A larger abundance of hydrogen in the envelope favour a redder star with a larger radius. 
Models M, N and O isolate the effect of the surface abundance of hydrogen for the same hydrogen gradient.
Again, the models with a larger abundance of hydrogen in the envelope favour a redder star. These models present another example of how certain definitions of the hydrogen gradient do not always fully capture the effects on \teff.
Models P, Q \& R mimic a possible effect of mass accretion or a merger with material of a different average H/He abundance, in which only the outer half of the envelope contains a different hydrogen abundance.
A similar effect of \xenv on \teff is found.

While the effect of \xenv will be intuitive to many readers, we can also understand it in terms of how it affects \lnuc and \lactual. An increase in the hydrogen abundance in the envelope (e.g. compare Models N and O) increases the opacity in the envelope, reducing the ability to transport energy by radiation and decreasing \lactual. A decrease in \lactual favours an increase in the stellar radius and a decrease in \teff, which is exactly what we see in the models. Although the impact of an increased hydrogen abundance in the envelope on \teff due to opacity effects has been understood for quite some time, our snapshot models clearly isolate the effect of \xenv.

\subsection{Helium Abundance in the Core} \label{sec:ycore}
\begin{figure} \centering
\centering
\includegraphics[width=\linewidth]{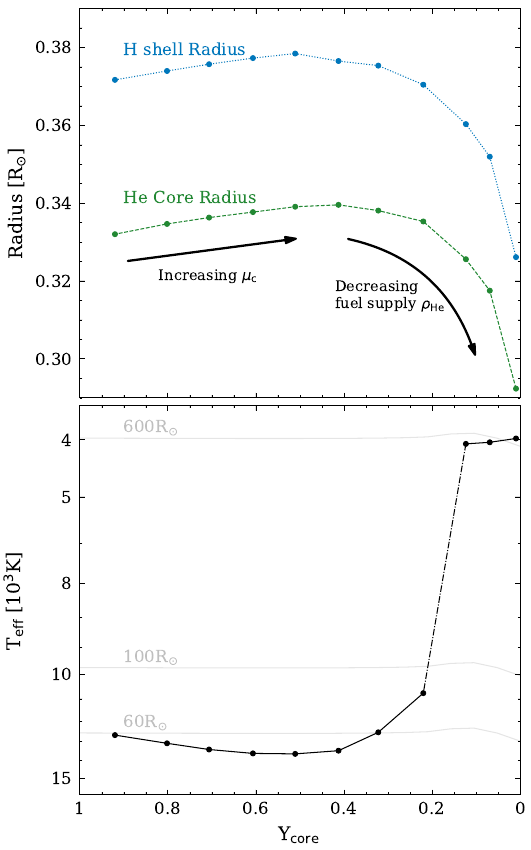}
\caption{\textit{Top Panel:} Effect of the central helium abundance \yc on the radius of the helium core, where $X < 10^{-4}$, and the radius of the peak of the H-burning shell for a set of snapshot models with the same core mass, envelope mass and abundance profile in the envelope (S-13). \textit{Bottom Panel:} Effect of \yc on \teff and the stellar radius for the same models as in the top panel.}
\label{fig:yc_teff}
\end{figure}

During the core helium burning phase, the helium abundance in the core, \yc, is primarily modified by nuclear burning. Its evolution as a function of time can also be affected by processes such as convective core overshooting, rotational mixing and, in evolutionary models, the $^{12}C(\alpha, \gamma)^{16}O$ reaction rate. Using the \textsc{snapshot} approach, we can isolate the effect of \yc on \teff and the stellar radius for a set of models with identical abundance profiles in the rest of the star outside the convective core Fig.~\ref{fig:yc_teff}. To support our discussion, we also plot the impact of \yc on the radius of the helium core in green, defined as where $X < 10^{-4}$, and the radius of the peak of the hydrogen burning shell in blue.

We find that \yc affects \teff in two competing ways. First, a decrease in \yc causes an increase in the mean molecular weight of the core \mucore as helium is converted to C and O. 
The increase in \mucore favours an increase in the radius of the helium core (Fig.~\ref{fig:yc_teff}). 
This is for the same reason that the radius of a core hydrogen burning star increases with increasing \mucore. 
A higher \mucore favours a higher temperature and density through the equation of state, increasing \lnuc and therefore favouring a larger helium core radius. 
However, a larger helium core radius favours a smaller total stellar radius. 
This is because as the core radius increases, so does the radius of the hydrogen burning shell, as it is located just above the helium core. 
As the radius of the hydrogen shell increases, the temperature and density throughout the burning region decrease, which leads to a decrease in \lnuc and favours a smaller radius. 
The second impact of \yc on \teff is through the decrease in the available fuel supply in the core for the triple alpha reaction. 
Again, similar to core hydrogen burning stars, a decrease in the fuel supply causes a decrease in \lnuc and favours a smaller helium core radius. 
A smaller helium core radius moves the radius of the hydrogen burning shell inwards, causing an increase in the temperature, density and \lnuc, favouring a larger total stellar radius. 
In summary, as \yc decreases, the increase in \mucore initially favours a decrease in the stellar radius while the decrease in the available fuel supply in the core subsequently favours an increase in the stellar radius.

For our 15 \msun representative model, the impact of \yc on \mucore dominates for $1 > \yc \gtrsim 0.50$ while the impact on the fuel supply dominates for $0.50 \gtrsim \yc > 0$.
Therefore, the effect of \yc on the surface properties in non-monotonic. 
For $1 > \yc \gtrsim 0.50$ a decrease in \yc favours an increase in \teff, while for $0.50 \gtrsim \yc > 0$ a decrease in \yc favours a decrease in \teff. 
We find that the quantitative impact of \yc on \teff is smaller for large values of \yc, i.e. the beginning of helium burning, and larger towards the end of helium burning for $\yc < 0.40$. 
The impact of the fuel supply on \teff in core hydrogen burning stars only dominates over the effect of the mean molecular weight for very low values of $\xc \lesssim 0.05$.
However, for core helium burning stars it can dominate for $\yc \lesssim 0.50$. 
This is likely related to a combination of the larger relative change in mean molecular weight when converting from $^{1}$H to $^{4}$He and from $^{4}$He to $^{12}$C as well as the smaller dependence of the CNO cycle reaction rate on the density of hydrogen compared to the dependence of the triple alpha reaction on the density of He.

We tested our interpretation of the effect of \yc on \teff by computing a similar set of snapshot models to those in Fig.~\ref{fig:yc_teff} but in which we keep the mean molecular weight constant, and just decrease the fuel supply with decreasing \yc. This is achieved using a similar method as for the core hydrogen burning models described in Sec.~\ref{ms_results}. In these models, we found that the radius of the core and H-shell decreased monotonically with decreasing \yc and the total stellar radius increased monotonically. This implies that the effect of a decreased fuel supply is indeed to favour an overall increase of the stellar radius in a core helium burning star, supporting our analysis above. 

Our snapshot models show the non-monotonic behaviour of $R_{\rm c}$ and $R$ with \yc.
In the context of blue loops in the HR diagram \citet{Lauterborn1971} discussed the impact of the radius of the helium core on \teff in terms of the parameter $\Phi_{\rm c} = M_{\rm c}/R_{\rm c}$, the ratio of the mass of the core to the radius of the core. 
As found by \citet{Lauterborn1971}, our models show that when $R_{\rm c}$ decreases, $R$ increases and vice versa. 
We also provide the main cause for these behaviors.

\subsection{Core Mass Ratio} \label{sec:mcoreratio}

\begin{figure*} \centering
\includegraphics[width=\linewidth]{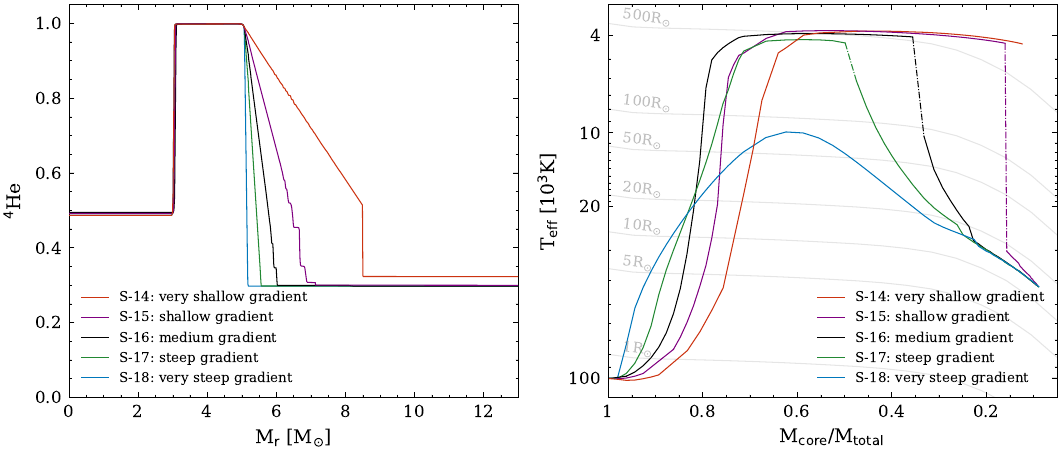}
\caption{\textit{Left panel:} The helium abundance profile in the hydrogen shell region for each sequence of models (S-14 to S-18). \textit{Right panel:} The effect of the combination of core mass ratio \coremassratio on the surface properties. The value of \teff as a function of core mass ratio is plotted for each sequence of models with the same core mass and varying envelope mass.}
\label{fig:menv_teff}
\end{figure*}

The combination of the core mass \mcore and envelope mass \menv of a star during helium burning can be affected by many different physical processes.
Core masses are intrinsically larger for higher mass stars due to the formation of larger convective core masses during the core hydrogen burning phase.
The core mass can also be increased by mixing during core hydrogen burning due to convective boundary mixing and rotation.
The envelope mass can decrease due to mass loss and can be dramatically modified by binary interaction via stripping, mass accretion or a merger.
In \citet{Farrell2020a}, we computed snapshot models to isolate the effect of the envelope mass on the surface properties for constant helium core mass and abundance.
Here, we extend this analysis by studying how the helium abundance in the hydrogen shell affects the relationship between the core mass, envelope mass and the surface properties.

We present five sequences of snapshot models, each with the same helium core mass and varying envelope mass. Each sequence has a different hydrogen/helium abundance profile in the hydrogen burning shell, as indicated in Fig. \ref{fig:menv_teff}.
For a given helium core mass, the envelope mass affects \teff in two competing ways.
On one hand, an increase in the mass of hydrogen in the envelope increases the effect of opacity.
This decreases the amount of energy that can be transported, causing a decrease in \lactual and favouring a larger radius. 
On the other hand, an increase in the mass of the envelope modifies the hydrostatic structure of the star and requires a larger value of \lactual in the envelope to support a higher mass. 
The increase in \lactual favours a bluer star with a smaller radius. 
For all of the model sequences in Fig. \ref{fig:menv_teff}, the core mass ratio has a non-monotonic effect on \teff. 
For core mass ratios $\coremassratio \gtrsim 0.6$, the effect of opacity dominates and therefore \teff decreases with increasing envelope mass (decreasing \coremassratio).
For core mass ratios $\coremassratio \lesssim 0.6$, the effect of increasing mass on the hydrostatic structure dominates and \teff increases with increasing \menv (increasing \coremassratio). 
The fact that massive stars during the post-main sequence with higher envelope masses tend to favour a blue supergiant solution rather than a red supergiant solution has been known in the literature for a long time.
However, a simple explanation for this has not always been clear.

Fig. \ref{fig:menv_teff} also clearly illustrates how the helium profile in the hydrogen burning shell affects the relationship between the core mass ratio and the surface properties.
In the stripped star regime (i.e. low envelope masses), at the same core mass ratio a larger abundance of helium in the hydrogen shell favours a bluer star.
This is simply due to the fact that the envelope mass is so small that the helium lowers the opacity of the envelope, favouring a smaller radius.
For intermediate and higher envelope masses, a high \yshell at the same core mass ratio favours a redder star.
In these models, the effect of the helium abundance on the H-shell energy generation dominates, increasing \lnuc and favouring a redder star as discussed in Sec. \ref{sec:yshell}.

At this point, the reader may wonder why it is that if you continue to add mass onto a typical core helium burning star, the radius decreases, but if you add mass to a typical main sequence star, the radius increases?
It appears that this is due to a sort of a boundary condition effect.
The boundary at the inner edge of the hydrogen burning shell in stars during helium burning is set by the hydrostatic and thermodynamic properties of the helium core.
However, the boundary at the center of main sequence stars is not constrained in the same way.
This allows the temperature and density to increase with increasing mass, increasing \lnuc and favouring a larger radius.

\subsection{Metallicity: CNO in the hydrogen shell and Z in the envelope} \label{sec:cnoshell}

\begin{figure*} \centering
\centering
\includegraphics[width=\linewidth]{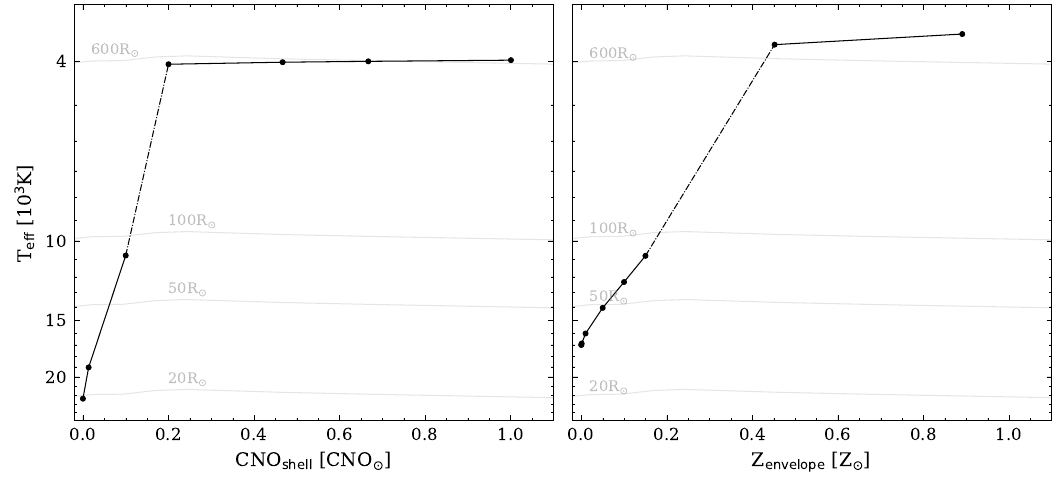}
\caption{\textit{Left panel:} Effect of the CNO abundance in the hydrogen burning shell on \teff for a representative set of stellar models with the same core mass, envelope mass, core composition and helium abundance profile (S-19). \textit{Right panel:} Effect of the metallicity in the envelope above the hydrogen burning shell (S-20).}
\label{fig:cnoteff}
\end{figure*}

Stellar evolution models show that core helium burning stars with a lower initial metallicity tend to have a higher \teff \citep[e.g.][]{Stothers1968}.
The initial metallicity of a star can impact its evolution in several ways including by modifying its mass loss rate, the efficiency of rotational mixing, the internal temperature structure and the convective core mass.
All of these evolutionary effects have complex feedback effects.
Because of this, it is difficult to use evolutionary models to isolate the effect of metallicity on post-MS stellar structures. 
Our snapshot models indicate that both the CNO abundance in the hydrogen burning shell (\cnoshell) and the metal opacity in the envelope have important impacts on \teff. 
The value of \cnoshell is predominantly determined by the initial metallicity but can be modified by rotational mixing if C and O are mixed from the core into the H-shell. 
This occurs most dramatically so in the case of low or zero metallicity stars \citep[e.g.][]{Ekstrom2008}.

Using snapshot models, we isolate the effect of \cnoshell on \teff for a representative 15 \msun model, in which we scale the abundance of all the CNO elements in the H-shell, while keeping the rest of the internal abundance profiles the same (Fig.~\ref{fig:cnoteff}). The CNO elements are converted to hydrogen to conserve mass. An increase in \cnoshell favours an increase in the stellar radius and a lower \teff. This can be understood by considering that a higher \cnoshell causes a higher \lnuc (due to the effect on the CNO cycle), favouring a redder star. 
We also test the effect of the abundance of the metals in the non-burning region of the envelope, above the hydrogen shell (Fig.~\ref{fig:cnoteff}). As expected, a higher abundance of metals in the envelope increases the effect of opacity, decreases \lactual and favours a larger radius.
The internal abundance profiles for these models are included in the online supplementary material.

Many previous works have studied and discussed the evolutionary effects of metallicity on \teff of core helium burning stars \citep[e.g][]{Schaller1992, Langer1995}. It has been pointed out by \citet{Stothers1968} that the \teff of massive post-main sequence stellar models may be lowered by increasing only the CNO abundances in the star. This can be connected to the effect of \cnoshell. 
Additionally, \citet{Schaller1992} found that a lower initial metallicity favours a more extended blue loops in the HR diagram during helium burning. 
This can also be understood by the fact that a lower initial metallicity implies a lower CNO content in the shell and lower metal content in the envelope which, as explained above, favour a more compact, bluer star.

\subsection{Comparison of effects in the HR diagram} \label{sec:hrd_heburning}

\begin{figure*} \centering
\includegraphics[width=1.0\linewidth]{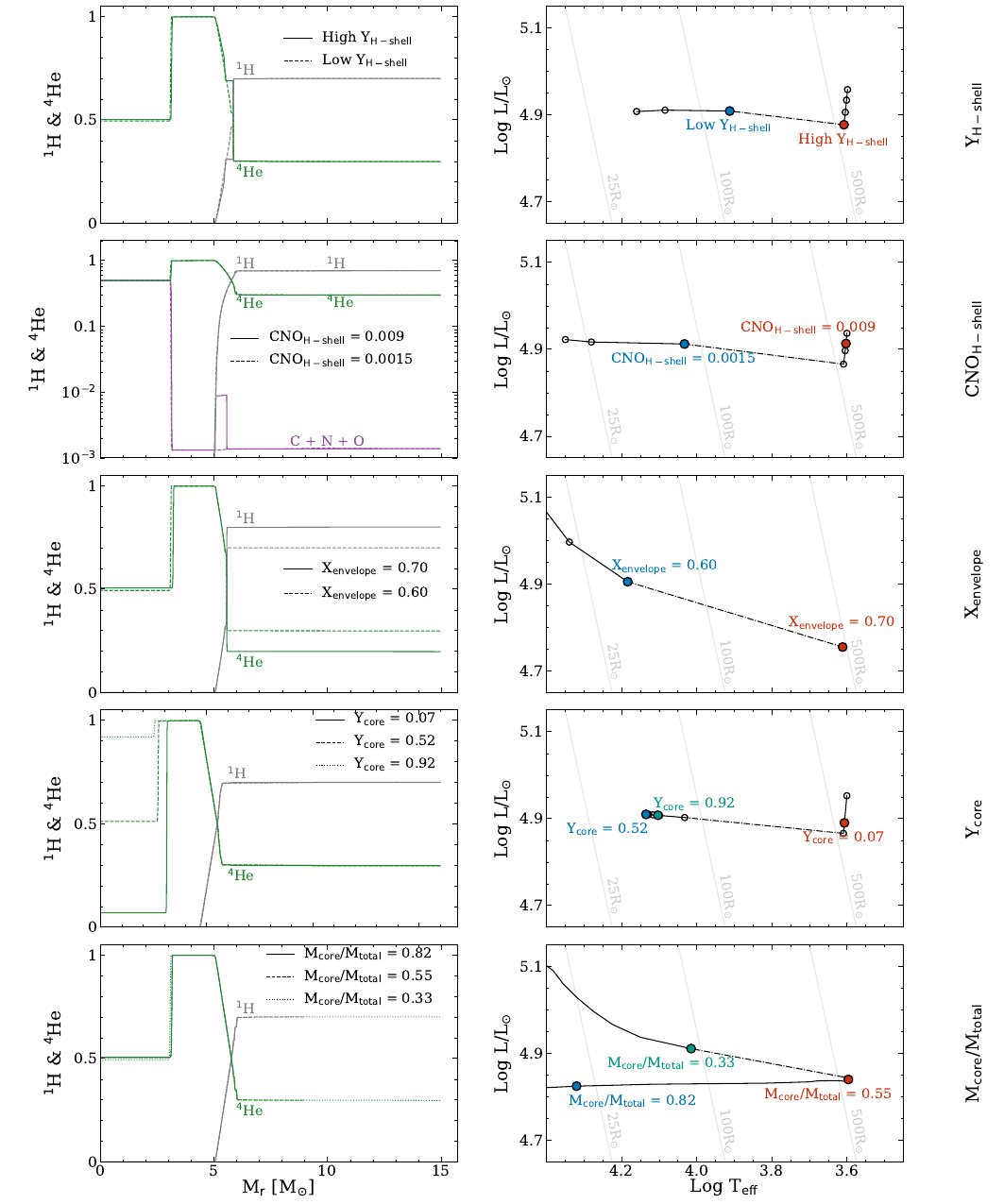}
\caption{Isolating the effect of key features of the internal abundance profile on the location of a 15 \msun star in the HR diagram. In each case, we plot the internal abundance profiles of hydrogen and helium for two or three stellar models in which only one property changes and the rest of the star remains the same. We also plot the location in the HR diagram of each of these models (circles) with a black line joining intermediate models (not plotted in the upper panels).  The dash-dot line indicates the bi-stability transition between a BSG and RSG.}
\label{fig:summary_6panel}
\end{figure*}

Fig. \ref{fig:summary_6panel} summarises and compares the effects of the five key features of the internal abundance profile discussed above on the values of \ly and \teff. For each property, we select representative models which demonstrate the important effects. The internal hydrogen and helium abundance profiles of two or three representative \snap models are plotted in the left panels and their location in the HR diagram are plotted in the right panels (blue, red and green circles), as well as several intermediate models that are not included in the abundance profiles (empty circles). The dash-dot lines in the HR diagram indicate the approximate location of the bi-stability transition between a blue and a red supergiant, in which intermediate stellar models in hydrostatic and thermal equilibrium do not exist. As these results relate only to the internal structure of stars in hydrostatic and thermal equilibrium, they are not affected by the prior evolution. Therefore, they apply to all stars regardless of the mass loss history, any internal mixing or binary interaction. Fig. \ref{fig:summary_6panel} is also relevant for the crossing of the HR diagram between the core hydrogen and helium burning phases, as we will discuss in Sec.~\ref{sec:hr_gap}.

In summary, Fig. \ref{fig:summary_6panel} demonstrates the following:
First, a larger helium abundance profile in the region of the H-burning shell, \yshell, favours a larger radius and lower \teff. 
In some cases, the stellar radius can change by a factor of 4 for a change in the helium mass of only of 0.05 \msun. \yshell does not modify the luminosity significantly except when the star is on the Hayashi track. 
As \yshell increases, the stellar radius increases, the envelope cools and more of the envelope becomes convective. 
A higher convective mass in the envelope increases the rate at which energy can be transported, increasing \lactual, resulting a higher surface luminosity. 
Second, the CNO abundance in the shell, \cnoshell, has a similar effect on \ly and \teff to \yshell, for similar reasons. 
An increase of \cnoshell by a factor of 6 decreases \teff from $11\,000$K to $4\,000$K. 
Third, an increase in \xenv also favours a lower luminosity due to the fact that an increase in the opacity causes a decrease in \lactual, resulting in a lower surface luminosity. 
A decrease of the hydrogen abundance in the envelope (\xenv) from 0.70 to 0.60 can cause a very large increase in \teff i.e. from $4\,000$K to $20\,000$K. 
Fourth, the effect of \yc on \teff is non-monotonic but most important during the second half of the core helium burning phase when it favours a decrease in \teff with decreasing \yc. The effect on the luminosity is similar to the previous properties, for the same reasons. 
Finally, the core mass ratio \coremassratio also has a non-monotonic effect on \teff. 
For \coremassratio decreasing from 1 to $\sim 0.6$ (i.e. corresponding to increasing envelope mass), the radius increases and \teff decreases. 
For further decreasing \coremassratio (i.e. further increasing envelope mass), the radius decreases and \teff increases. 
The luminosity gradually increases with increasing envelope mass due to the larger luminosity produced/required by the hydrogen burning shell.

\section{The Crossing of the HR diagram after the main sequence} \label{sec:hr_gap}

\begin{figure} \centering
\includegraphics[width=\linewidth]{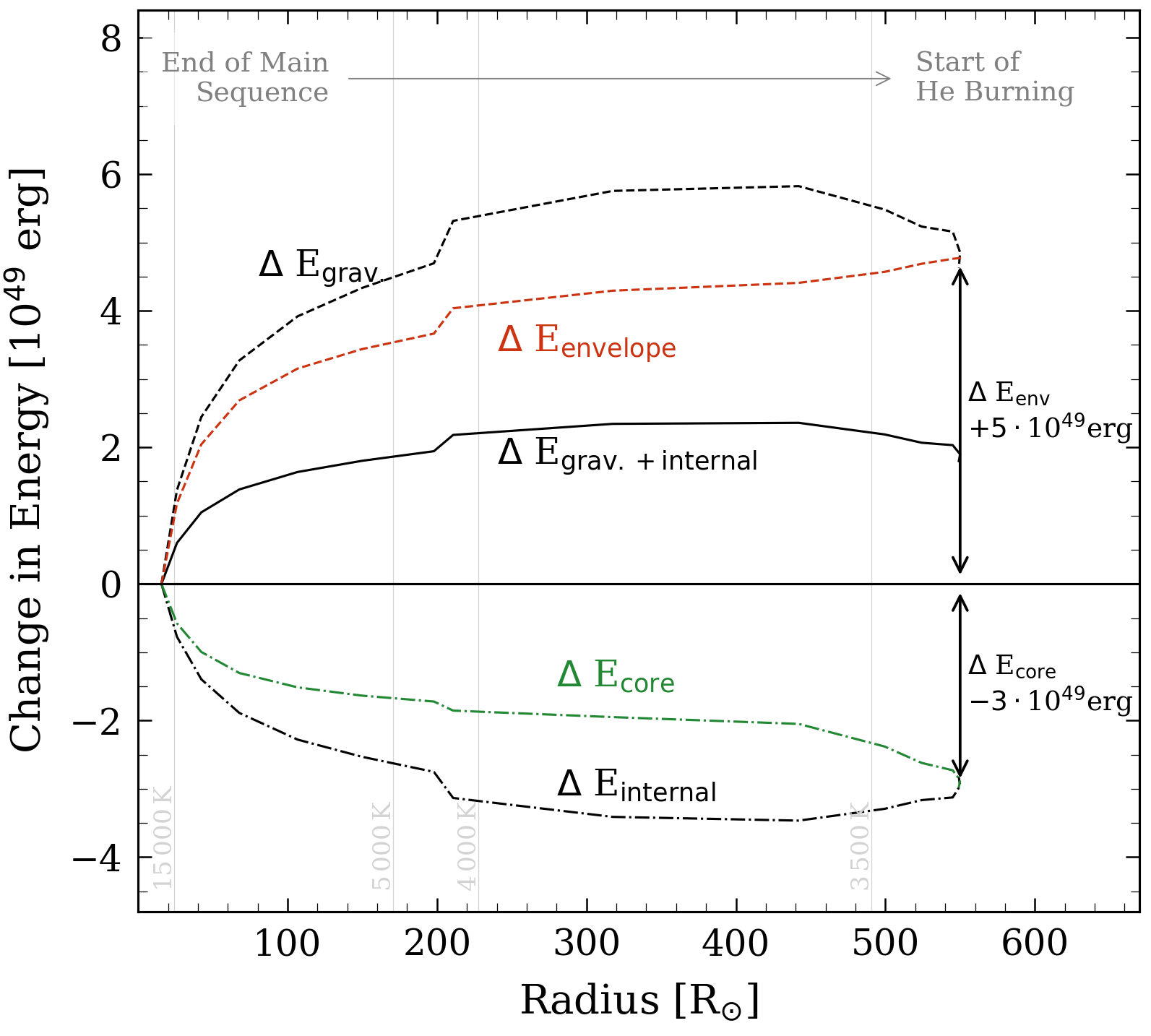}
\caption{The change in internal energy (dash-dot), gravitational energy (dashed) and total energy (solid) of a representative 12 \msun model as it expands from the end of the main sequence to the beginning of core helium burning (T-21). Also included are the total gravitational + internal energy of the core (green) and the envelope (red).}
\label{fig:hrgap_energies}
\end{figure}

\begin{figure} \centering
\includegraphics[width=\linewidth]{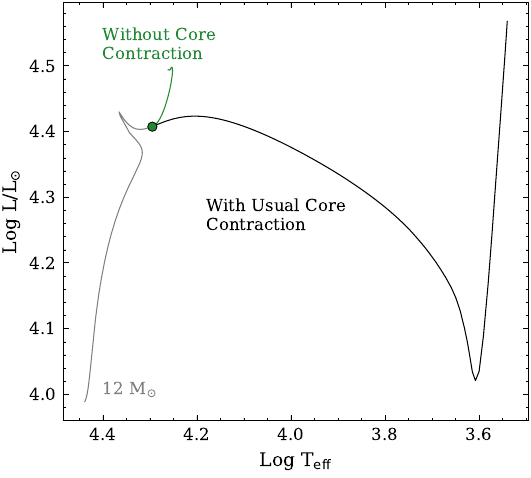}
\caption{Effect on the surface properties if we artificially suppress the contraction of the core during the crossing of the HR diagram after core hydrogen exhaustion (green line, T-22) compared to the usual expansion (black line) in a representative 12 \msun star at solar metallicity (T-21).}
\label{fig:hrgap_stopcore}
\end{figure}

\begin{figure} \centering
\includegraphics[width=\linewidth]{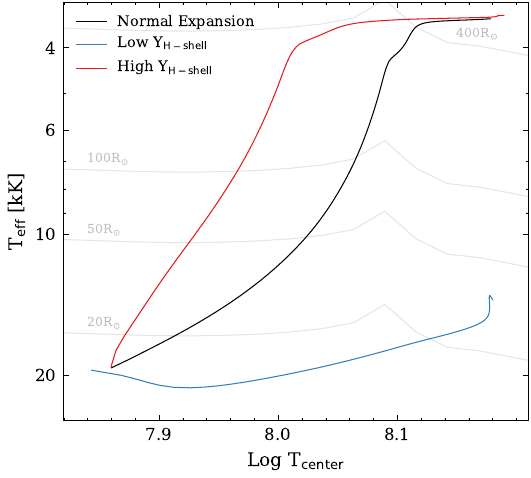}
\caption{The effect of the helium abundance in the H-burning shell \yshell on the expansion of a 12 \msun star from the end of the MS to the start of core helium burning (T-23). The usual stellar evolution model is shown in black as well as models with a higher (red) and lower (blue) value of \yshell.}
\label{fig:hrgap_yshell}
\end{figure}

\begin{figure} \centering
\includegraphics[width=\linewidth]{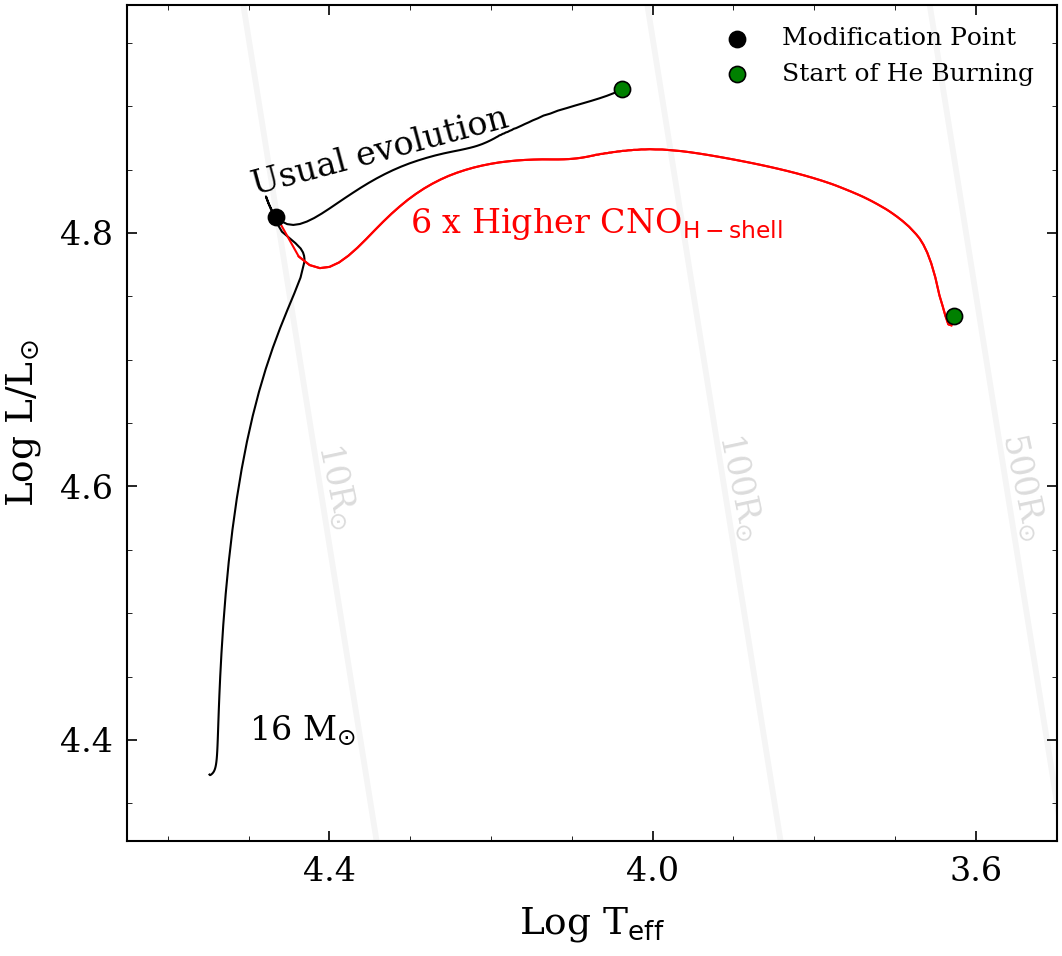}
\caption{Effect of the CNO abundance in the H-burning shell \cnoshell on the expansion across the HR diagram after core hydrogen exhaustion in a 16 \msun model at a metallicity of Z = 0.002 (T-24). The usual evolution to the beginning of core helium burning in shown in black. The red line shows the expansion when we artificially increase the CNO abundances in the hydrogen shell at the black point by a factor of 6.}
\label{fig:hrgap_cnotest}
\end{figure}

When hydrogen is exhausted in the core at the end of the main sequence phase, most stars expand significantly to become giants or supergiants. 
A simple explanation for why this happens currently appears somewhat elusive, despite extensive discussion in the literature \citep{Hoppner1973, Eggleton1981, Yahil1985, Applegate1988, Eggleton1991, Renzini1992, Iben1993, Sugimoto2000, Stancliffe2009, Ball2012}. 
To investigate this expansion in massive stars, we perform numerical tests on representative 12~\msun stellar models and apply our understanding of what sets \ly and \teff from Sec. \ref{theory}, \ref{ms_results}, \ref{he_results}. 
One point we would like to emphasise is that any compelling explanation for why stars expand after the main sequence should also describe why some stars (or stellar models) expand more than others. 
For example, comparing Fig.~\ref{fig:hrd_evolutions1} and Fig.~\ref{fig:hrd_evolutions2} the 16~\msun stellar model expands to about 80 \rsun at SMC metallicity (Z = 0.002), compared to 600 \rsun at a metallicity ten times larger (Z = 0.020).

When the central hydrogen abundance decreases below about $\xc = 0.05$ in our standard (unmodified) 12~\msun stellar evolution model, the entire star contracts due to the decrease in the central fuel supply, decreasing the total (gravitational + internal) energy of the star. The region above the hydrogen depleted core heats up and begins to burn hydrogen, forming the hydrogen burning shell. 
The hydrogen burning shell prevents that region from contracting due to the stabilising feedback of nuclear burning \citep{Maeder2009a}. 
However, the  hydrogen depleted core continues to contract as it has no other way to support itself. 
The contraction of the core has two effects. 
Firstly, it moves the radial position of the  hydrogen shell inwards, increasing the temperature and density of the hydrogen shell and, hence, increasing \lnuc. 
As discussed in Sec.~\ref{theory}, this favours evolution to a larger radius and a lower \teff. 
Secondly, it transfers some energy from the core (due to the decrease in gravitational potential energy) to the envelope. 
At any given point during the expansion, the envelope can only transport a certain amount of energy. 
This excess energy cannot be transported by the envelope, so it cools and expands. 

In our  $12\,\msun$ model, 60\% of the increase of the energy of the envelope is accounted for by the decrease in the energy of the core and 40\% is accounted for by increased nuclear energy generation in the hydrogen shell as a result of the contraction. 
Of course, these fractions will likely be different for models of different masses.
A consequence of this is that the effect of a given amount of core contraction on the total stellar radius will depend on other features of the internal abundance profile including \yshell, \cnoshell, \xenv and \coremassratio. 
Any property of the envelope that favours a larger radius when the whole star is in thermal equilibrium will also cause the star to expand by more for a given amount of contraction by the core (i.e. a given change in the central temperature or radius of the core). 

To test this explanation, we perform three tests using modified stellar evolution models. 
Fig.~\ref{fig:hrgap_stopcore} shows the effect on the surface properties if we artificially suppress the contraction of the core during the crossing of the HR diagram after core hydrogen exhaustion (green line) compared to the usual expansion (black line) in a representative 12 \msun star at solar metallicity. 
The evolution during the main sequence phase is plotted in grey.
Just after core hydrogen exhaustion (green circle), we effectively ``remove the core'' by inserting fixed inner boundary conditions equal to the values at the boundary of the core. 
For this purpose, we define the core as the central region that is contracting. 
Maintaining fixed inner boundary conditions, we follow the response of the star in the HR diagram, indicated by the green line in Fig. \ref{fig:hrgap_stopcore}. 
The star does not expand, remaining in the blue region of the HR diagram. 
The luminosity increases slightly due to the position of the hydrogen shell artificially moving slightly inwards. 
However, the star does not expand. 
This supports the understanding that if the hydrogen shell is not continually forced to contract and heat up by the core, the star will simply not expand. 
We also perform a test (not plotted) in which we turn off the hydrogen shell burning. In this test, the star does not expand, but rather stays in the blue region of the HR diagram and contracts. 
The lack of the stabilising feedback provided by nuclear burning in the  hydrogen shell means that the whole star continues to contract.

We now investigate how the abundance profile in the  hydrogen shell (\yshell from Sec. \ref{sec:yshell}) affects the expansion across the HR diagram. 
Along with the standard $12\,\msun$ evolutionary model from Fig. \ref{fig:hrgap_stopcore}, we compute two models with larger and smaller amounts of helium in the hydrogen burning shell.
In Fig.~\ref{fig:hrgap_yshell}, we plot the value of \teff, representing the expansion of the envelope across the HR diagram, as a function of the central temperature, representing the contraction of the core.
In all three models, the timescale of the expansion is the same. 
The expansion of the star proceeds on the Kelvin-Helmholtz timescale of the core. As the core mass is the same for each model, the timescale of the expansion is also the same for each model.
The expansion of the standard, unmodified model is plotted in black. 
The high \yshell model with a higher abundance of helium in the hydrogen shell expands at a faster rate as a function of the contraction of the core than the standard, unmodified model (in black). 
Conversely, the low \yshell model expands at a slower rate than the standard model and actually begins core helium burning as a blue rather than a red supergiant. 
This can be understood by considering that a given contraction of the core will cause a given increase in the temperature and density in the hydrogen shell. 
However, the quantitative effect on \lnuc, and therefore on the stellar radius, depends on the properties of the hydrogen shell. 
As discussed in Sec. \ref{sec:yshell}, a larger value of \yshell, i.e. a shallower hydrogen gradient, favours a redder star. 

We perform a similar investigation for the abundance of the CNO elements in the  hydrogen shell \cnoshell (Fig.~\ref{fig:hrgap_cnotest}). 
We begin with a $16\,\msun$ stellar model with a metallicity of Z = 0.002.
Just after core hydrogen exhaustion (black point), we increase the abundance of the CNO elements just in the region of the  hydrogen shell (similar to Fig.~\ref{fig:summary_6panel}) by a factor of 6. 
We then observe the response of the model as it expands to begin core helium burning (red line). 
It expands to much larger radii and lower \teff than the original model, beginning helium burning as a red supergiant rather than a blue supergiant. 
This can be understood in a similar way to the effect on \yshell. 
Stellar evolution models of low metallicity stars tend to favour beginning core helium burning as blue supergiants, compared to higher metallicity models which favour core-He ignition as a RSG. 
This is not because their cores are hotter so they take less time to contract to become hot enough to ignite helium and therefore have less time to expand. 
Rather it is the effect of the lower CNO abundance in the hydrogen burning shell, combined with the lower metal opacity in the envelope.
In fact, there is kind of a coincidence here: the core is hotter because there is a lower CNO abundance in the core and the star begins helium burning at a higher \teff because there is a lower CNO abundance in the hydrogen shell. 
Of course, the lower CNO abundances are both due to the lower metallicity.

\section{Cause and Effect in the HR Diagram} \label{cause_effect_summary}
In this section, we apply our results from Sections \ref{ms_results}, \ref{he_results} and \ref{sec:hr_gap} to explain which features of the internal abundance profiles dominate the change in \ly and \teff at different points in the evolution of a star. We select three representative cases at intermediate and high masses: a 16 \msun star at solar metallicity (Z = 0.020), a 16 \msun star at SMC metallicity (Z = 0.002) and a 6 \msun star at solar metallicity.

\subsection{The Evolution of a 16 Solar Mass Star at Z = 0.020}

\begin{figure} \centering
\includegraphics[scale=1.0]{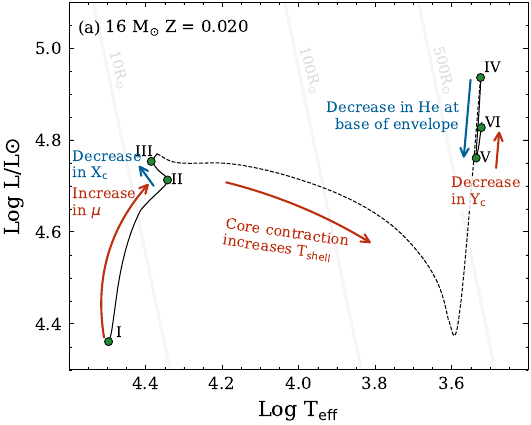}
\caption{The evolution of a 16 \msun star at solar metallicity (Z = 0.020) in the HR diagram, with arrows and text indicating the primary cause for the evolution in each direction. The dashed line indicates the transition from the MS to core helium burning. The following six stages of the evolution are highlighted: I is at the zero-age main-sequence, II is at the terminal-age main-sequence when the star begins to contract, III is when the star begins to expand and cross the HR diagram, IV is the beginning of core helium burning, V is when \yc = 0.30 and VI is at the end of core helium burning.}
\label{fig:hrd_evolutions1}
\end{figure}

Point I in Fig.~\ref{fig:hrd_evolutions1} indicates the zero-age main sequence for a $16 \msun$ star at solar metallicity (Z = 0.020).
Our \textsc{snapshot} models (Fig. \ref{fig:ms_hrd}) recover the well-known results for the MS \citep[e.g.][]{Kippenhahn1990, Maeder2009}.
As hydrogen is converted to helium via the CNO cycle during the MS phase, the average mean molecular weight increases which causes an increase in the surface luminosity and the stellar radius.
When the central hydrogen mass fraction \xc drops below about $\xc = 0.05$ (point II), the entire star contracts, causing the star to evolve to the left in the HR diagram (the Henyey hook).
This is caused by the decrease in the central fuel supply.
The luminosity continues to increase due to the additional energy released from the gravitational contraction in the outer layers of the star.
The increase in temperature just above the hydrogen-depleted core due to the contraction creates a hydrogen burning shell (point III).

Due to the feedback effect of nuclear burning \citep{Maeder2009a}, the H-burning shell acts to limit any further contraction (or expansion) in its vicinity. 
The core continues to contract, which has two main effects (Sec. \ref{sec:hr_gap}). 
Firstly, it converts the core's gravitational potential energy partly to internal energy and partly to the luminosity which supports the core (Fig. \ref{fig:hrgap_energies}). 
Secondly, it changes the hydrostatic structure of the star, increasing the temperature and density at the base of the hydrogen burning shell.
This increases the rate of nuclear energy generation in the shell. 
The extra energy produced in the shell cannot be transported by the envelope in a timescale shorted than the core contraction timescale, so the envelope cools and expands.
The cooling and expanding of the envelope is reflected in the increase of the stellar radius from $10 \rsun$ to $500 \rsun$ (from points III to IV).
This expansion proceeds on the Kelvin-Helmholtz (KH) timescale of the core.

When the central temperature and density are high enough to burn helium via the triple-alpha reaction, the core stops contracting (point IV). 
This stops the increase in temperature and density at the H-burning shell.
As a result, the envelope stops expanding and the star begins to evolve on a nuclear burning timescale again. 
At this point, the hydrogen and helium abundance profiles in the H-burning shell are defined mainly by the one left behind from the core hydrogen burning phase. 
As the hydrogen shell burns through this profile, the abundance profiles of hydrogen and helium in the burning region change. 
The amount of helium in the shell initially decreases, causing a decrease in the stellar radius and the star evolves back down the Hayashi track to higher \teff (Sec. \ref{sec:yshell}).
At point V, the evolution in the HR diagram reverses and the star evolves back toward larger radii and lower \teff, due to the the decrease in central helium abundance (Sec. \ref{sec:ycore}). 
At point VI, the star depletes its core helium.

\subsection{The Evolution of a 16 Solar Mass Star at Z = 0.002}
\begin{figure} \centering
\includegraphics[scale=1.0]{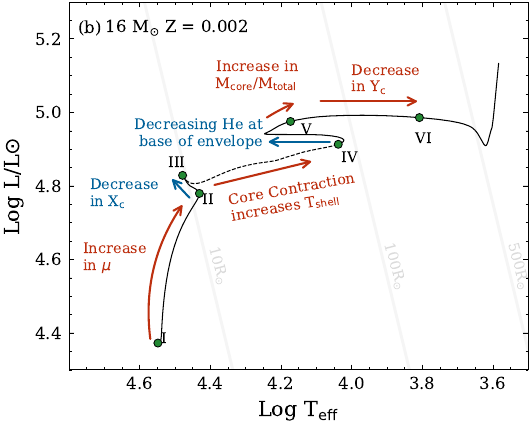}
\caption{Same as Fig. \ref{fig:hrd_evolutions1}, but for a 16 \msun star at SMC metallicity (Z = 0.002).}
\label{fig:hrd_evolutions2}
\end{figure}

A $16\,\msun$ star at SMC metallicity (Z = 0.002) begins its evolution with a smaller radius at a higher \teff and a higher luminosity than at solar metallicity (Fig.~\ref{fig:hrd_evolutions2}, point I) . 
A lower abundance of CNO elements in the center of the star and of metals in the envelope results in a smaller radius that at solar metallicity (Sec. \ref{sec:cnoshell}). 
The higher luminosity is a result of higher $T$ and $\rho$ in the outer parts of the nuclear burning region.
The evolution during the MS is the similar to the solar metallicity model until core hydrogen depletion (point III). 
After the main sequence, the rate of expansion of the envelope with the contraction of the core is smaller than for the solar metallicity model. 
Again, this is due to lower CNO abundances in the H-burning shell and metals in the envelope (Sec. \ref{sec:cnoshell}).

Due to the lower CNO abundance in the H-burning shell, the low metallicity model begins burning helium as a BSG (point IV). Similar to the solar metallicity model, the H-shell burns through the profile left behind from the MS and the expansion after the MS, changing the amount of helium in the H-shell, \yshell. The envelope contracts and the star evolves to higher \teff as \yshell decreases (Sec. \ref{sec:yshell}). Once \yshell remains relatively constant, the contraction stops and the star begins to evolve back towards the red region of the HR diagram. The increase in \coremassratio as a result of the growth of the mass of the core drives the evolution to lower \teff (Sec. \ref{sec:mcoreratio}). Subsequently, the decrease in \yc dominates, which also drives evolution to lower \teff (Sec. \ref{sec:ycore}) until it becomes a RSG. Our models indicate that a massive star at solar metallicity (Z = 0.020) may expand to become a RSG for a different reason than a star of the same mass at SMC metallicity (Z = 0.002).

\subsection{The Evolution of a 6 Solar Mass Star at Z = 0.020}
\begin{figure} \centering
\includegraphics[scale=1.0]{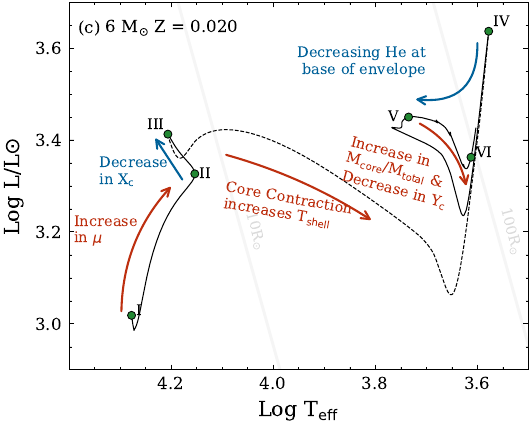}
\caption{Same as Fig. \ref{fig:hrd_evolutions1}, but for a 6 \msun star at solar metallicity (Z = 0.020).}
\label{fig:hrd_evolutions3}
\end{figure}

An intermediate mass $6\,\msun$ star evolves similarly to the $16\,\msun$ model beginning of helium burning (Fig.~\ref{fig:hrd_evolutions3}). 
Once core helium burning begins, the hydrogen shell burns through the profile above the core driving the star to lower radii and higher \teff, as for $16\,\msun$ model (Sec. \ref{sec:yshell}). 
However in the $6\,\msun$ case, due to the lower core mass ratio \coremassratio, the star evolves to much lower radii than the $16\,\msun$ model. 
It evolves away from the Hayashi line and toward the blue region of the HR diagram and the beginning of a blue loop is formed. 
The model spends about 70\% of its helium burning lifetime as a BSG at higher \teff. 
The combination of the subsequent increase in \coremassratio and decrease in \yc makes the star evolve back to the Hayashi line (Sections \ref{sec:mcoreratio} and \ref{sec:ycore}).

\section{Other Discussion Points} \label{discussion}

\subsection{Blue Loops in the Hertzsprung-Russell Diagram}
Stellar evolution models in the mass range 5 - 12 \msun sometimes exhibit a “blue loop”, in which a star evolves from a RSG to the blue region of the HR diagram and back to the red, completing a loop in the HR diagram (e.g. Fig.~\ref{fig:hrd_evolutions3}). 
Their existence has been known for a long time \citep{Hayashi1962, Hofmeister1964} and they have been extensively discussed in the literature \citep{Schlesinger1977, Stothers1979, Maeder1981a, Walmswell2015}. The properties of blue loops are known to be highly sensitive to processes such as convective overshooting, mass loss and semi-convection.
They are especially important in the context of the production of Cepheids. 
In the following paragraph, we use the model from Fig. \ref{fig:hrd_evolutions3} and our results from Sec. \ref{he_results} to explain why blue loops occur.

As a RSG evolves, its helium core increases in mass as hydrogen burns in a thin shell around the core.
As it burns outwards, the helium abundance profile in the hydrogen shell may change, which can affect the stellar radius (Sec. \ref{sec:yshell}).
In typical stellar evolution models at these masses, the change in the helium profile during helium burning favours a bluer star.
At the beginning of core helium burning, the helium profile in the shell favours a large radius due to its high helium content.
As the shell moves outwards through the helium profile, the profile in the burning region changes to favour a blue star.
If this effect outweighs the other effects of the core mass ratio, affected by e.g. convective overshooting, mass loss and rotation, and the CNO abundance in the hydrogen shell, then the star evolves back towards the blue and may become a BSG.
After the initial decrease in radius, the combined effect of decreasing \yc and increasing core mass ratio due to hydrogen shell burning favour evolution to a larger radius.
At some point, this effect wins out and the star evolves back to the red. 
Our results suggest that blue loops in intermediate and massive stars should be favoured at relatively lower masses because of their intrinsic lower core mass ratios. 
This is consistent with what we find in stellar evolution models and in observations of Cepheids.
Our models also suggest that processes that increase the core mass ratio such as increased convective overshooting or post-MS mass loss will disfavour the production or extent of blue loops, which is also consistent with previous studies.

The “mirror effect” has been invoked as a phenomenological description of the blue loops \citep{Hayashi1962, Hofmeister1964, Sandage1952}.
In this description the expansion of the core causes a contraction of the envelope as the cause of the star evolving back to the blue, in the opposite way to a star expanding across the HR gap.
Our results indicate that the slight expansion of the edge of the core is an effect of the changing helium profile above the core star, rather than a cause of the blue loop.

\subsection{Why do stars tend to become more luminous and expand as they evolve?}

The temperature and density in the nuclear energy generation regions increase as a star evolves. 
This is primarily due to either an increased higher mean molecular weight in the burning region or a change in the hydrostatic structure due to the contraction of the core when it runs out of fuel. 
The increased temperature and density cause an increase in \lnuc. 
As a consequence, stars usually produce slightly more energy than they can transport, causing a cooling and expansion of the envelope. 
The luminosity goes up if the increase in \lnuc outweighs the effect of the expansion on the burning region. 
This is always the case during MS and sometimes the case during post-MS. 
Additionally, the evolution of most of the key features of the internal abundance profile e.g. \coremassratio, \yc, \yshell favour a larger radius as the star becomes more evolved. 
This means that the more a star evolves, the harder it is to maintain a small radius.

\section{Conclusions} \label{sec:summary}

In this paper, we aimed to isolate the key features of the internal abundance profile that drive stellar evolution during different evolutionary stages. We summarise the key findings below.

\begin{enumerate}
    \item We devised a framework to qualitatively understand cause and effect in the evolution of the surface properties of stars that is ultimately based on an argument from conservation of energy.
    We discussed how changes in \lnuc, the cumulative internal luminosity distribution generated by nuclear reactions, and \lactual, the actual internal luminosity distribution, can help to provide an qualitative understanding for why a star evolves to a given \ly and \teff.
    Beginning in thermal equilibrium, any change to the internal abundance profile, the hydrostatic structure or the energy transport that causes an increase in \lnuc or a decrease in \lactual will favour evolution to a larger radius, and vice versa.
    
    \item We isolated and quantified the key features of the internal abundance profile that set the surface properties for stars during the main sequence, the core helium burning phase and the short-lived expansion in between. Our results provide a new way to interpret observations of individual stars and stellar populations in terms of the structural properties that favour a given set of observed properties.
    
    \item Massive stars with lower metallicity tend to have higher \teff for two reasons: (i) lower CNO abundances in the core (hydrogen burning) and the H-shell (helium burning) which affects nuclear energy generation and (ii) lower opacity in the envelope. During the post-main sequence, the effect of the CNO abundances dominates for BSGs while the effect of opacity dominates for RSGs.
    
    \item Models of massive main sequence stars with the same mass and very similar surface properties can have different internal distributions of hydrogen and different convective core masses. This degeneracy might be broken with current and future asteroseismology observations.
    
    \item Massive stars expand after the main sequence because the contraction of the core heats up the hydrogen-burning shell, generating more energy than the envelope can transport in a typical core contraction timescale. Whether a star begins helium burning as a blue or red supergiant depends on the helium and CNO abundances in the hydrogen shell, the core mass ratio and opacity due to hydrogen and metals in the outer envelope. Each of these properties affect the rate of expansion of the envelope for a given contraction of the core.
    
    \item We discuss the cause of blue loops in the HR diagram during the post-main sequence. Consistent with previous works, we find that the key factor is the shape of helium profile in the hydrogen burning shell as the shell moves outwards during core helium burning. We also discuss the cause for why other factors, including the core mass ratio and metallicity, can cause stellar evolution models to exhibit blue loops. This has important implications for interpreting observations of Cepheids.
    
    \item We present a numerical test that clearly demonstrates that small changes in the stellar interior can cause very large changes in the surface properties.
    We conclude that much of the sensitivity seen in post-main sequence massive star models is ultimately due to the strong dependence of the CNO cycle energy generation rate on temperature in the hydrogen burning shell, which can be modified by very small changes in the helium abundance profile. 
\end{enumerate}

Consistent with previous work, our results show that a careful analysis of the internal abundance profiles is important for understanding how stars evolve.
Given current uncertainties in the internal mixing of massive stars, it is possible that the abundance profiles in the current state-of-the-art stellar evolution models differ substantially from actual stars. 
This may have significant impacts for our overall picture of stellar evolution, including studies of binary interaction and gravitational wave progenitors.

\section*{Data availability} The derived data generated in this research will be shared on reasonable request to the corresponding author.

\bibliographystyle{mnras}
\bibliography{main_refs, mesa_refs}

\section{Software Details} \label{sec:software}

Calculations were done with \textsc{MESA} version 15140.
The \textsc{MESA} EOS is a blend of the OPAL \citep{Rogers2002}, SCVH
\citep{Saumon1995}, FreeEOS \citep{Irwin2004}, HELM \citep{Timmes2000},
and PC \citep{Potekhin2010} EOSes.
Radiative opacities are primarily from OPAL \citep{Iglesias1993,
Iglesias1996}, with low-temperature data from \citet{Ferguson2005}
and the high-temperature, Compton-scattering dominated regime by
\citet{Buchler1976}.  Electron conduction opacities are from
\citet{Cassisi2007}.
Nuclear reaction rates are from JINA REACLIB \citep{Cyburt2010} plus
additional tabulated weak reaction rates \citet{Fuller1985, Oda1994,
Langanke2000}.  Screening is included via the prescription of \citet{Chugunov2007}.
Thermal neutrino loss rates are from \citet{Itoh1996}.

\section{Main Sequence Model Profiles} \label{sec:ms_model_profiles}

\begin{figure} \centering
\includegraphics[scale=1.0]{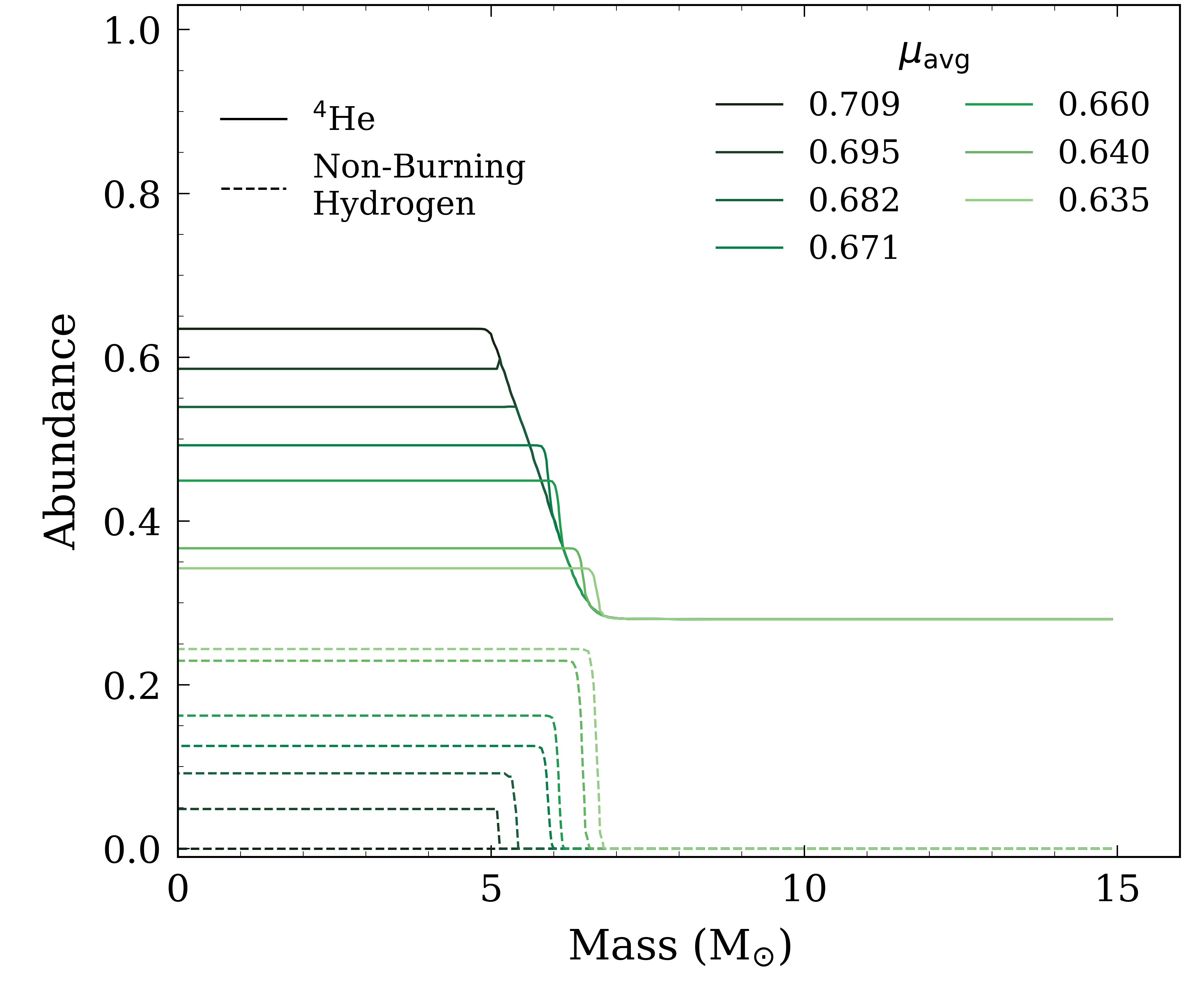}
\caption{Internal abundance profiles for main sequence models from Fig. \ref{fig:ms_hrd} in the mean molecular weight sequence. For the meaning of Non-burning hydrogen, see the discussion in Sec. \ref{ms_results}.}
\label{fig:mshrd_app1}
\end{figure}

\begin{figure} \centering
\includegraphics[scale=1.0]{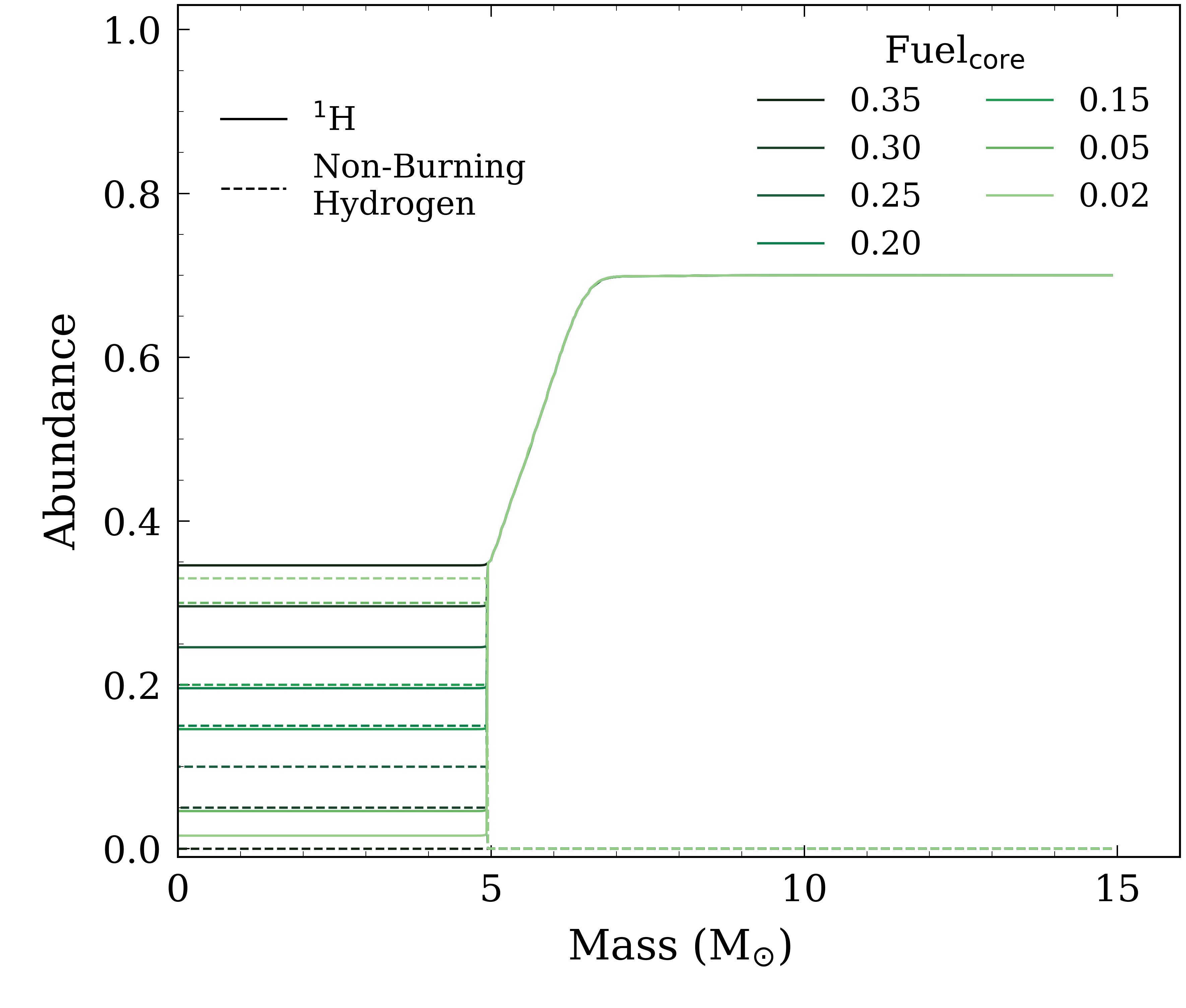}
\caption{Internal abundance profiles for main sequence models from Fig. \ref{fig:ms_hrd} in the fuel supply sequence. For the meaning of Non-burning hydrogen, see the discussion in Sec. \ref{ms_results}.}
\label{fig:mshrd_app2}
\end{figure}

\begin{figure} \centering
\includegraphics[scale=1.0]{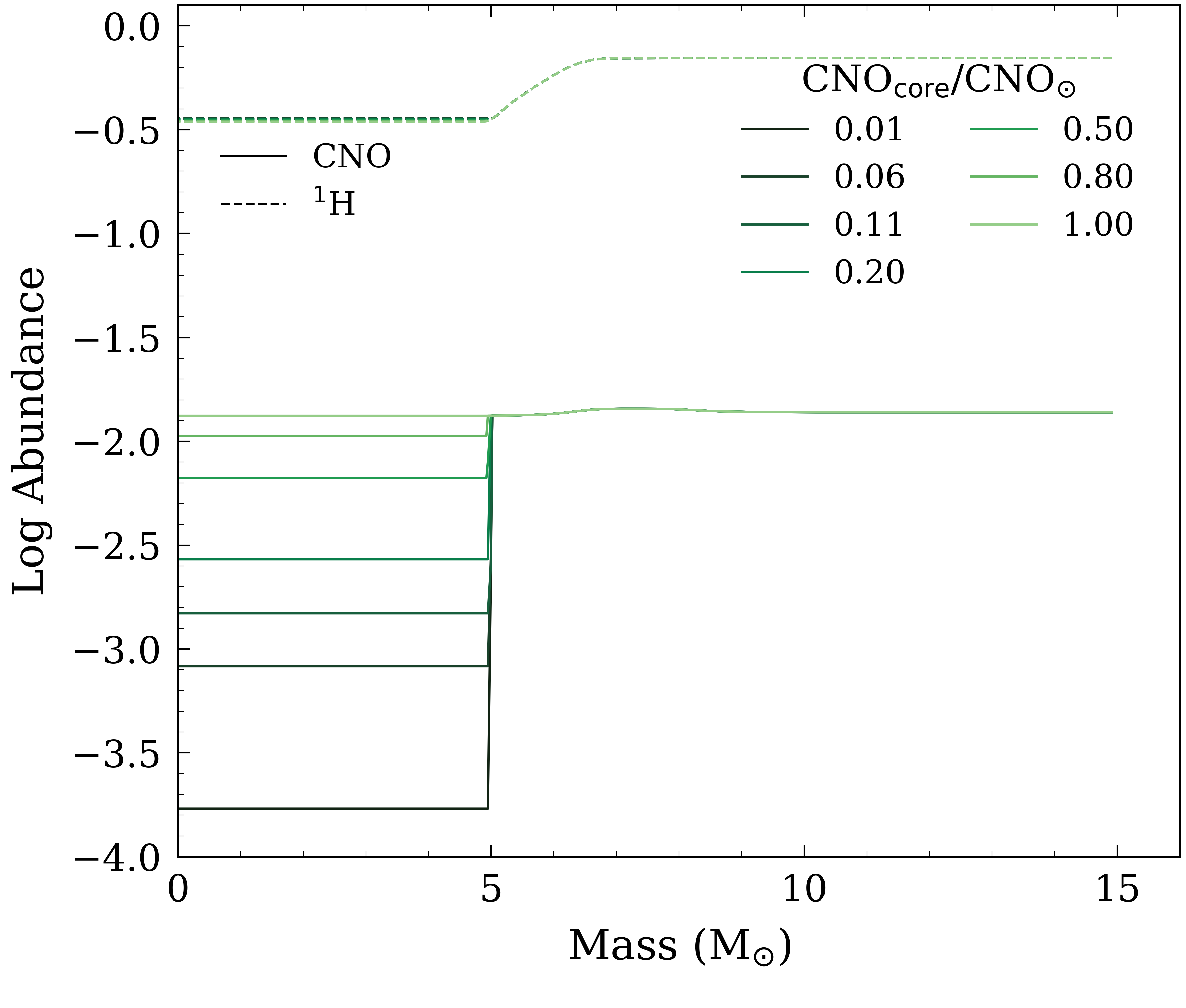}
\caption{Internal abundance profiles for main sequence models from Fig. \ref{fig:ms_hrd} in the CNO core abundance sequence.}
\label{fig:mshrd_app3}
\end{figure}

\begin{figure} \centering
\includegraphics[scale=1.0]{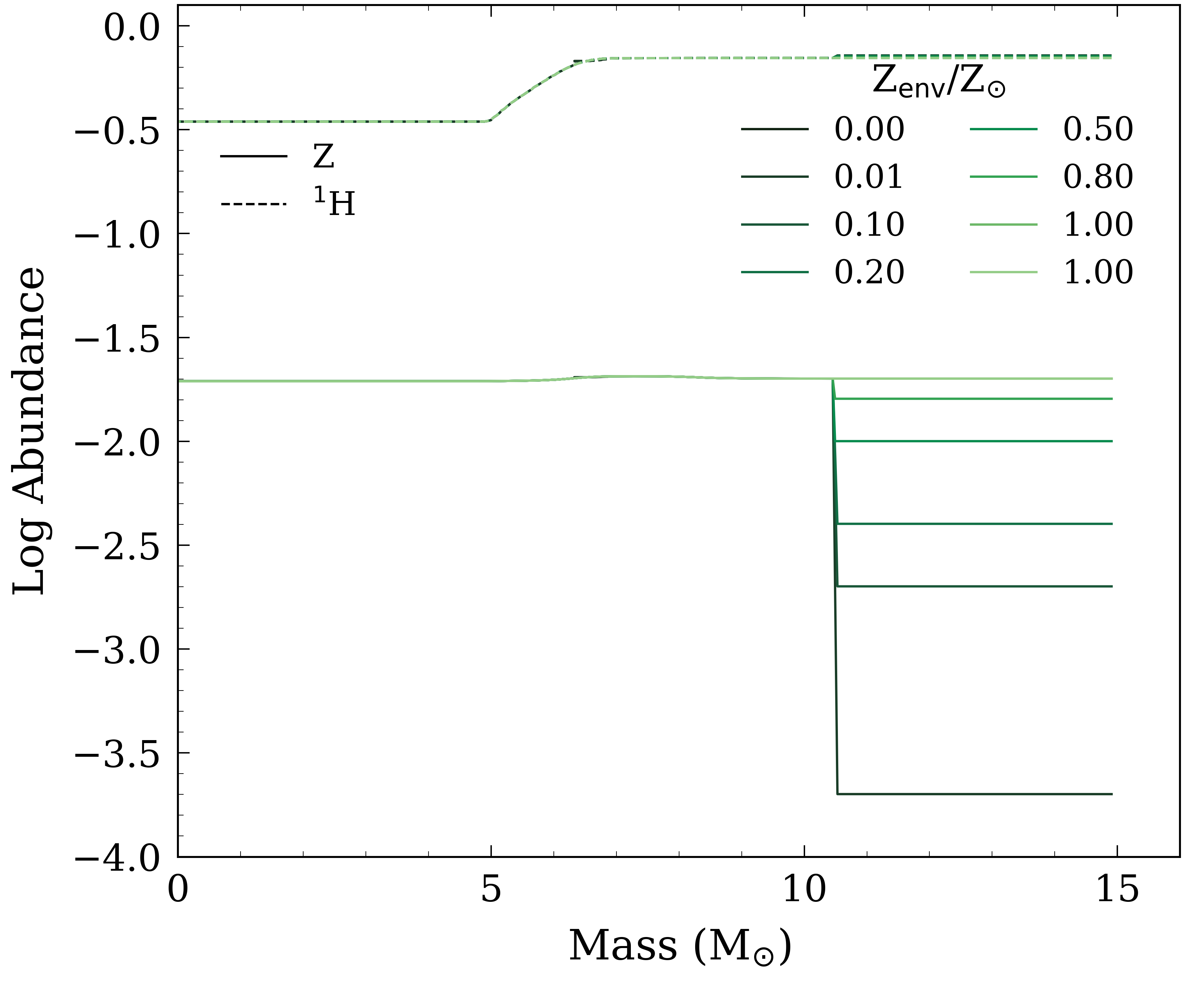}
\caption{Internal abundance profiles for main sequence models from Fig. \ref{fig:ms_hrd} in the envelope metallicity sequence.}
\label{fig:mshrd_app4}
\end{figure}

\begin{figure} \centering
\includegraphics[scale=1.0]{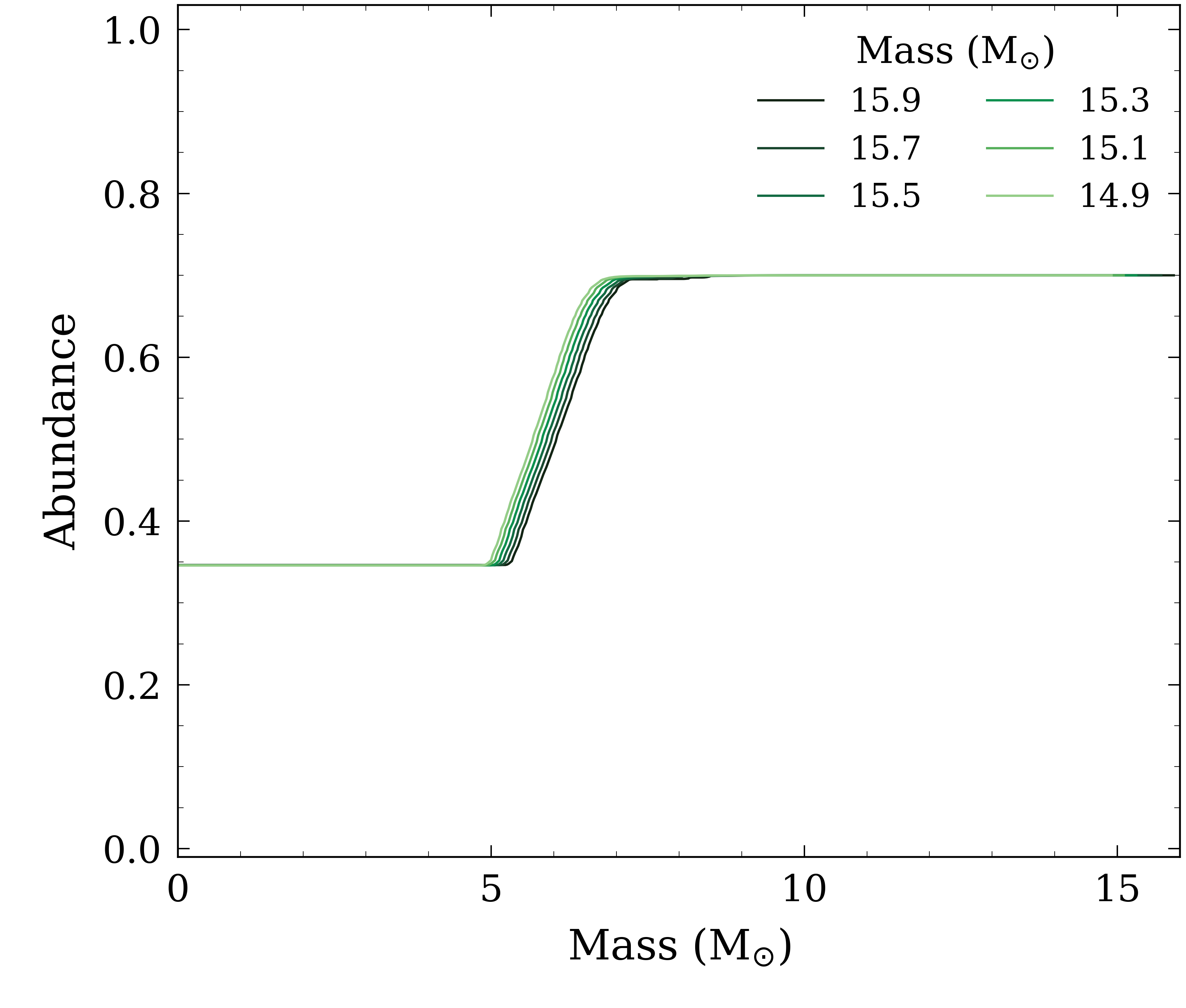}
\caption{Internal abundance profiles for main sequence models from Fig. \ref{fig:ms_hrd} in the total stellar mass sequence.}
\label{fig:mshrd_app5}
\end{figure}

\begin{figure} \centering
\includegraphics[scale=1.0]{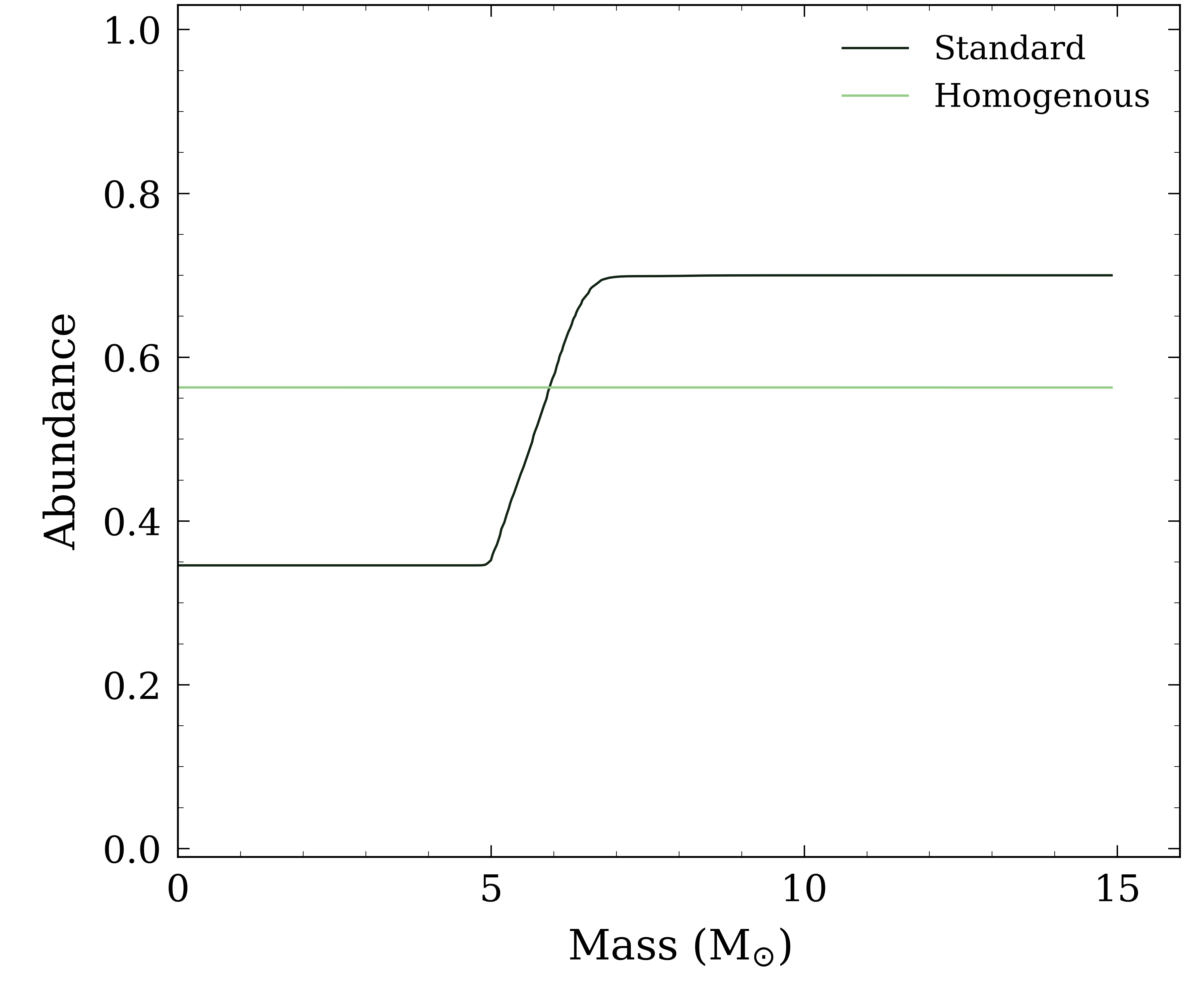}
\caption{Internal abundance profiles for main sequence models from Fig. \ref{fig:ms_hrd} in the hydrogen profile sequence.}
\label{fig:mshrd_app6}
\end{figure}

\section{Core Helium Burning Internal Profiles} \label{sec:heburn_model_profiles}

\begin{figure} \centering
\includegraphics[scale=1.0]{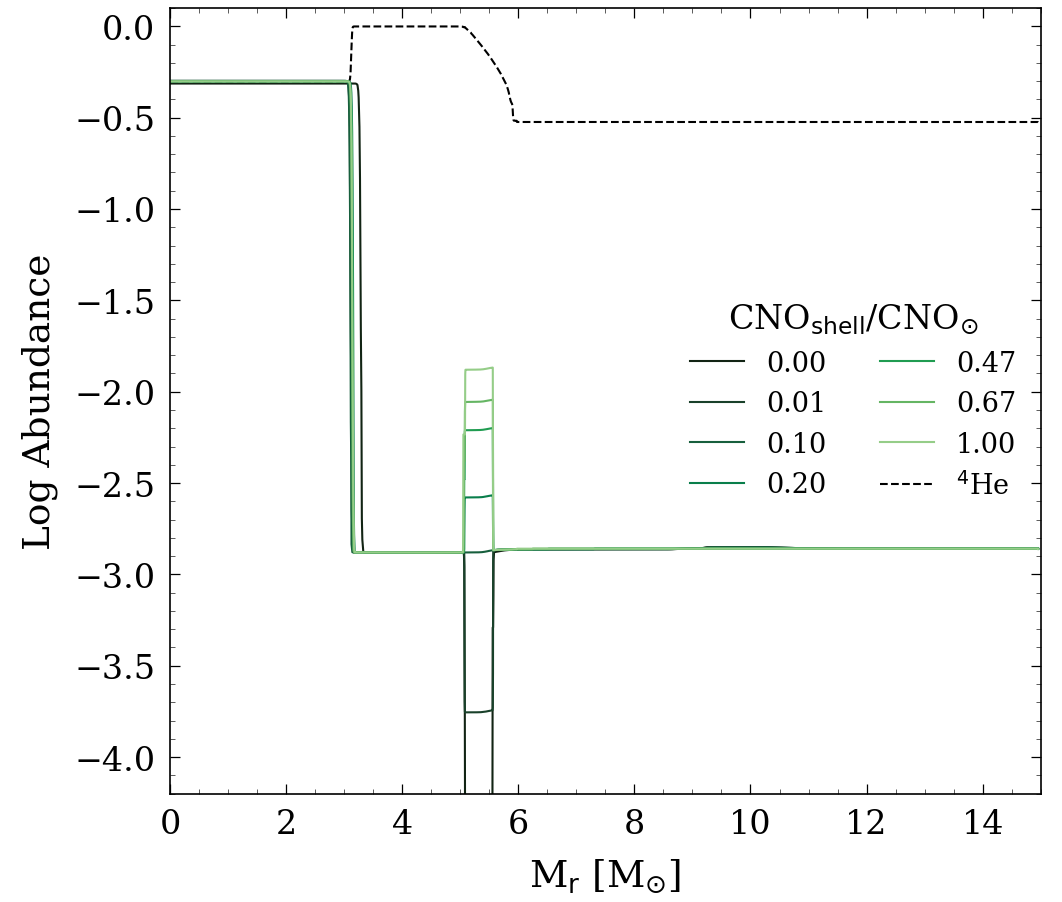}
\caption{Internal abundance profiles for helium burning models with varying CNO abundance in the hydrogen shell.}
\label{fig:cno_profile}
\end{figure}

\begin{figure} \centering
\includegraphics[scale=1.0]{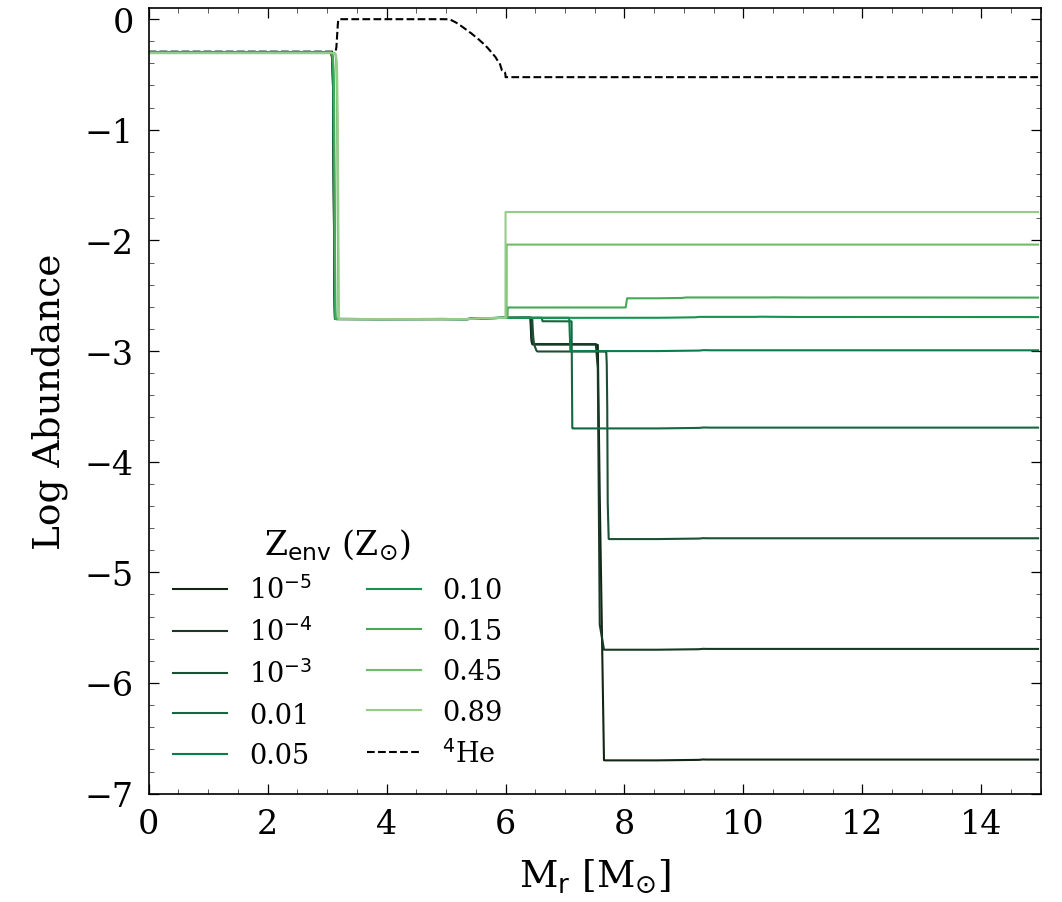}
\caption{Internal abundance profiles for helium burning models with varying metal abundance in the envelope.}
\label{fig:zenv_profile}
\end{figure}

\section{Detailed Numerical steps to support our conclusions}

\begin{figure*} \centering
\includegraphics[width=\linewidth]{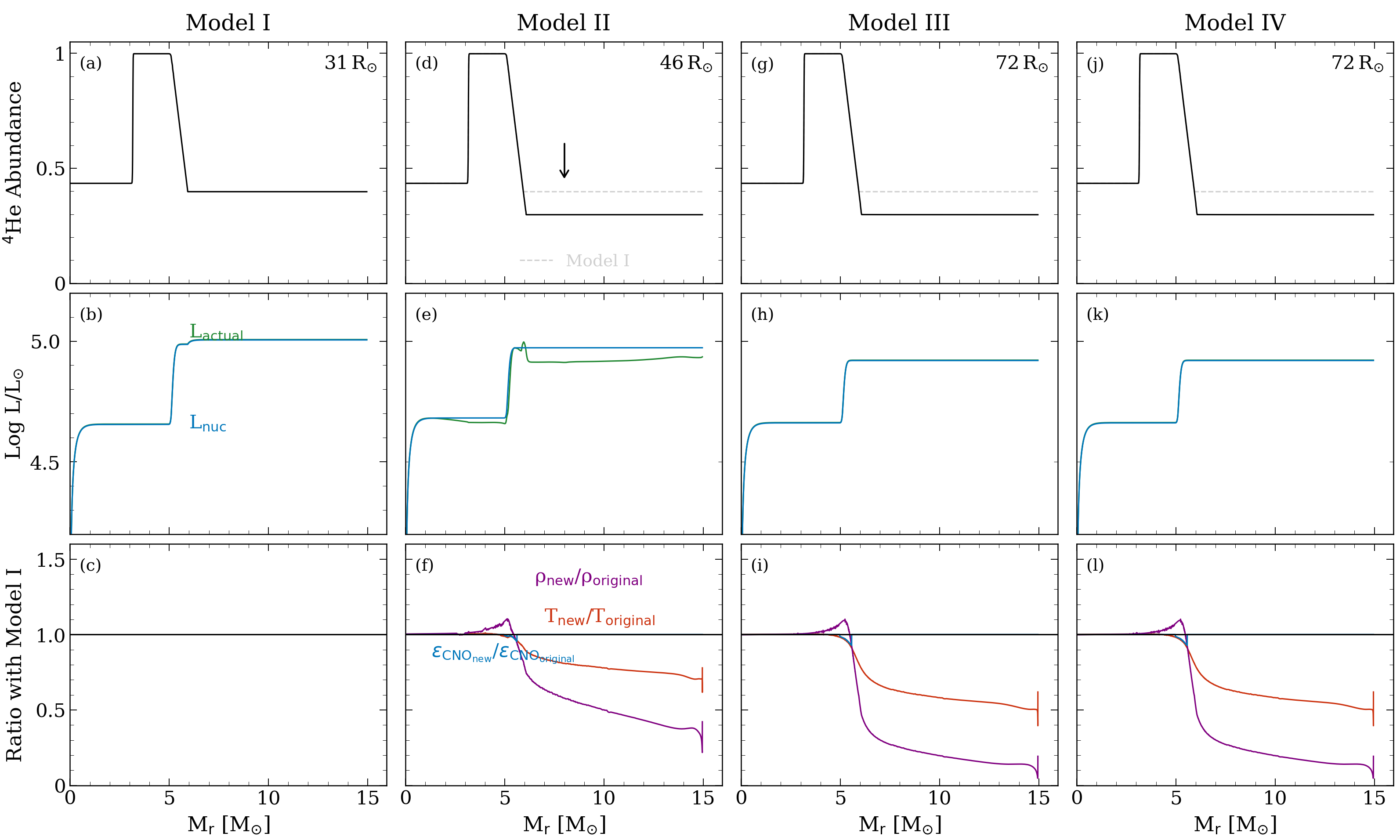}
\caption{A numerical test to examine the response of a star to changes in its hydrogen abundance profile in the envelope. \textit{Model I}: the original blue supergiant in thermal equilibrium. \textit{Model II}: stellar model immediately after the envelope abundance profile is modified, indicated by the black arrow in (d). \textit{Model III}: a few timesteps after Model II after the star has expanded to a radius of $72~\rsun$. \textit{Model IV}: when the star has reached thermal equilibrium as a red supergiant. For each model, we plot the internal profiles of the $^{4}$He, temperature (Log T scaled by a factor 0.2), the cumulative nuclear energy generation profile \lnuc, the luminosity imposed by hydrostatic equilibrium \lactual and the ratio with Model I of the internal temperature, density and $\epsilon_{\rm nuc}$ from CNO burning (scaled by a factor of 0.15) profiles.}
\label{fig:testa1}
\end{figure*}

\begin{figure*} \centering
\includegraphics[width=\linewidth]{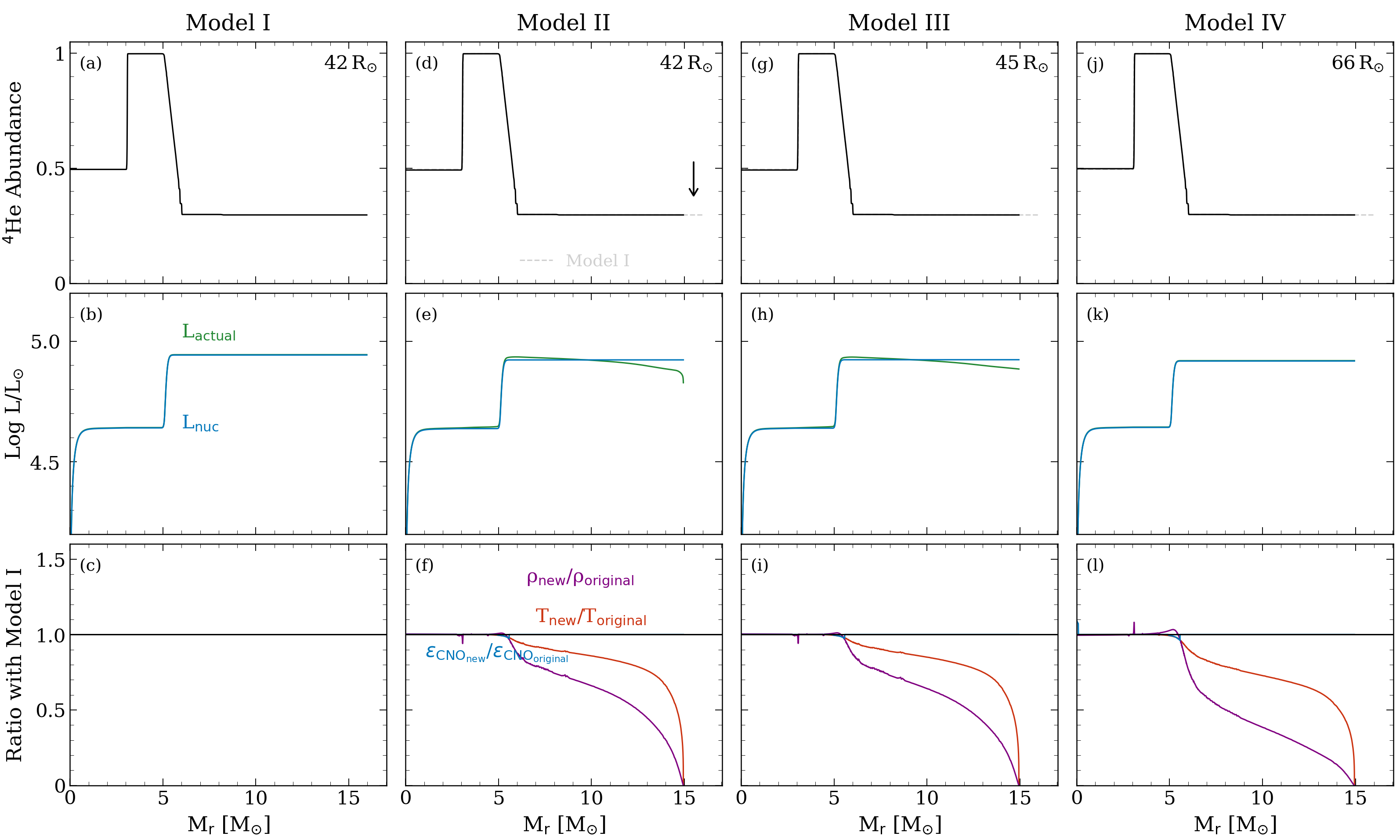}
\caption{A numerical test to examine the response of a star to changes in its envelope mass. \textit{Model I}: the original blue supergiant in thermal equilibrium. \textit{Model II}: stellar model immediately after the envelope mass is decreased, indicated by the black arrow in (d). \textit{Model III}: a few timesteps after Model II after the star has expanded to a radius of $45~\rsun$. \textit{Model IV}: when the star has reached thermal equilibrium as a red supergiant. For each model, we plot the internal profiles of the $^{4}$He, temperature (Log T scaled by a factor 0.2), the cumulative nuclear energy generation profile \lnuc, the luminosity imposed by hydrostatic equilibrium \lactual and the ratio with Model I of the internal temperature, density and $\epsilon_{\rm nuc}$ from CNO burning (scaled by a factor of 0.15) profiles.}
\label{fig:testa2}
\end{figure*}

\label{lastpage}
\end{document}